\documentclass[aps,prd,floatfix,10pt,nofootinbib,showpacs,notitlepage,superscriptaddress,longbibliography]{revtex4-1} 

\newcommand{\nn}{\nonumber \\}
\newcommand{\eq}[1]{Eq.~(\ref{#1})}
\newcommand{\fig}[1]{fig.~\ref{#1}}
\newcommand{\Fig}[1]{Fig.~\ref{#1}}

\newcommand{\bsub}{\begin{subequations}}
\newcommand{\esub}{\end{subequations}}
\newcommand{\eg}{\ensuremath{&=&}} 
\newcommand{\be}{\begin{eqnarray}}
\newcommand{\ee}{\end{eqnarray}}
\newcommand{\om}{\ensuremath{\omega}}

\newcommand{\pd}{\ensuremath{\partial}}
\newcommand{\la}{\ensuremath{\lambda}}
\newcommand{\ep}{\ensuremath{\epsilon}}
\newcommand{\lp}{\ensuremath{\left(}}
\newcommand{\rp}{\ensuremath{\right)}}
\newcommand{\abs}[1]{\ensuremath{\left \lvert #1 \right\rvert}} 
\newcommand{\bi} {\begin{itemize}}
\newcommand{\ei} {\end{itemize}}
\newcommand{\ben}{\begin{enumerate}}
\newcommand{\een}{\end{enumerate}}
\newcommand{\bmat}{\begin{pmatrix}}
\newcommand{\emat}{\end{pmatrix}}

\newcommand{\symbHS}{H_S} 
\newcommand{\symbHE}{M_H}  
\newcommand{\symbpiR}{\ensuremath{\pi r}} 

\usepackage[colorlinks=true,linkcolor=blue,citecolor=blue,urlcolor=blue]{hyperref}
\usepackage{graphicx}
\usepackage{amsmath}
\usepackage{slashed}  
\usepackage[latin1]{inputenc}   
\usepackage[english]{babel}
\usepackage{color}
\usepackage{amssymb}
\usepackage{mathrsfs}

\begin{document}
\title{Dynamical instabilities and quasi-normal modes, a spectral analysis with applications to black-hole physics} 

\author{Antonin Coutant}
\email{antonin.coutant@nottingham.ac.uk}
\affiliation{School of Mathematical Sciences, University of Nottingham, University Park, Nottingham, NG7 2RD, UK.}

\author{Florent Michel} 
\email{florent.michel@th.u-psud.fr}
\affiliation{Laboratoire de Physique Th\'eorique, CNRS, Univ. Paris-Sud, Universit\'e Paris-Saclay, 91405 Orsay, France.}

\author{Renaud Parentani}
\email{Renaud.Parentani@th.u-psud.fr}
\affiliation{Laboratoire de Physique Th\'eorique, CNRS, Univ. Paris-Sud, Universit\'e Paris-Saclay, 91405 Orsay, France.}

\date{\today}

\begin{abstract} 
Black hole dynamical instabilities have been mostly studied in specific models. We here study the general properties of the complex-frequency modes responsible for such instabilities, guided by the example of a charged scalar field in an electrostatic potential. We show that these modes are square integrable, have a vanishing conserved norm, and appear in mode doublets or quartets. We also study how they appear in the spectrum and how their complex frequencies subsequently evolve when varying some external parameter. When working on an infinite domain, they appear from the reservoir of quasi-normal modes obeying outgoing boundary conditions. This is illustrated by generalizing, in a non-positive definite Krein space, a solvable model (Friedrichs model) which originally describes the appearance of a resonance when coupling an isolated system to a mode continuum. In a finite spatial domain instead, they arise from the fusion of two real frequency modes with opposite norms, through a process that closely resembles avoided crossing. 
\end{abstract}


\maketitle

\section{Introduction}

The stability of black holes is an unexpectedly rich and complicated subject. Indeed many stationary black hole solutions develop dynamical instabilities in some regions of their parameter 
space~\cite{Press72,Damour76b,Gregory93,Kaloper,Nakamura:2009tf,Ooguri:2010kt,Gauntlett_helical_BHs,Brito:2015oca}. 
To understand in qualitative terms the origin of these instabilities, it is useful to consider the propagation of scalar fields in a four dimensional Kerr black hole geometry. When the field is massless, there is an \emph{energetic instability}, that is, the spectrum of stationary modes propagating in the region outside the event horizon contains negative energy solutions. The mixing of these modes with the usual asymptotic positive energy modes gives rise to a mode amplification, called in this context \emph{superradiance}~\cite{Zeldovich72,Unruh:1974bw,Starobinski73}. Importantly, this energetic instability can evolve into a {\it dynamical instability} when some reflection sends back the amplified modes towards the horizon. This results in a {\it pair} of modes with complex conjugated frequencies. Hence one of them exponentially grows in time. Efficient reflection can arise either from a non vanishing mass~\cite{Damour76b,Kaloper}, from the boundary of $AdS$~\cite{Press72,Cardoso:2004nk,Cardoso:2004hs,Cardoso:2006wa}, 
or even from a magnetic field~\cite{Brito14}. From these examples we learn that dynamical instabilities are closely related to super-radiant effects and energetic instabilities.

Besides these instabilities, the study of the resonances of stable black holes, usually called Quasi-Normal Modes (QNM), has also attracted a lot of attention~\cite{Kokotas99,Berti09,Konoplya11}. QNM are decaying solutions which obey outgoing boundary conditions. Interestingly, they are closely related to the modes responsible for dynamical instabilities (hereafter called DIM). As we shall see, this can be understood from the analytical properties of Green functions in the complex frequency plane
and from the fact that the DIM which exponentially grow in time also obey outgoing boundary conditions. 

Moreover, new types of instabilities are encountered when considering modified theories of gravity. In five dimensions, the infinite black string turns out to be dynamically unstable~\cite{Gregory93}. Indeed, the spectrum of gravitational perturbations obeying the outgoing boundary conditions contains a growing mode with a purely imaginary frequency. In $AdS_5$, when adding magnetic fields with a Chern-Simons coupling, above a certain threshold, the black hole horizon develops a helical instability~\cite{Nakamura:2009tf,Ooguri:2010kt}. In this case as well, the associated DIM is closely related to one QNM below the threshold. Higher dimensional black holes (branes) also possess 
some DIM~\cite{Gregory93,Cardoso:2005vk,Harmark:2007md,Gregory:2011kh}. In addition, analogue black holes~\cite{lrr-2011-3,PhysRevA.67.033602} 
display new types of instabilities. One interesting example is the black hole laser instability~\cite{Corley99,Coutant10} which is due to negative-energy modes bouncing back and forth between the two horizons. These closed trajectories are made possible because Lorentz invariance is broken by superluminal dispersive effects in the ultraviolet sector. 

At late time, non-linear effects will suppress the exponential growth of the unstable modes. The case of the Gregory-Laflamme instability was considered in~\cite{Park:2004zr, Lehner:2011wc}, where it was shown that the black string horizon fragments into an infinite number of black-hole horizons with spherical topology, arranged in a fractal structure. In \cite{Ooguri:2010kt,Nakamura:2009tf}, the instabilities of Reissner-Nordstr\"om black holes in 5D anti-de Sitter space coupled to the Maxwell-Chern-Simons theory, and its possible final states, were studied. It was found that unstable modes break spatial symmetries, leading to helical end-states. The saturation of the black hole laser effect in Bose-Einstein condensate, turning the region between the white and black hole horizons into a stable ``shadow soliton'' solution, or generating superposed infinite soliton trains, was discussed in \cite{MichelParentani,Michel:2015pra,deNova:2015fea}. The non-linear evolution of the charged black-hole bomb was studied in \cite{Sanchis-Gual:2015lje,Bosch:2016vcp}. In that case, there is a condensation of the scalar field extracting charge from the black hole. It is clear from these examples that non-linear effects are very model-dependent. They will not be considered in the present work, which focuses on generic features
of the solutions of linear wave equations.  

It is also clear from the above examples that instabilities are an essential part of black hole physics. So far they have mostly been studied in a case by case analysis. In the present article 
instead, we shall study DIM in general terms from a more abstract point of view. While resonances have been extensively studied in the context of the Schr\"odinger equation, the corresponding analysis of DIM has received much less attention, as was pointed out by Fulling in the Appendix of his book~\cite{Fulling}. The mathematical reason is that dynamical instabilities only arise if the conserved scalar product associated with the wave equation is non-positive definite. In such spaces, called ``Krein spaces''~\cite{Bognar} in the mathematical literature, no general spectral theorem is available. This contrasts with the case of positive definite Hilbert spaces~\cite{ReedSimon,LevyBruhl}, where general theorems play a crucial role for the spectral theory. 

Our first aim is to identify the key properties of DIM. To illustrate the various aspects, following~\cite{Fulling}, we first study the propagation of a two dimensional charged field in a square electrostatic potential. In this case, the possible frequencies are given by a transcendental equation which can be solved numerically. The evolution of QNM into DIM can thus be explicitly followed. As we shall see, this evolution is nontrivial and requires a careful study of the cuts of the wave vector in the complex frequency plane. Then, we shall show how the mode spectrum is related to the retarded and the advanced Green functions. This reveals the key role played by the asymptotic boundary conditions. When focusing on the mode spectrum, we shall see that the appropriate choice is that the modes are asymptotically bounded, and not that they obey the QNM outgoing condition. 

Our second aim is to generalize the well-known fact that an excited state of an isolated system (e.g. an atom) becomes a resonance, i.e., a pole in the second Riemann sheet of the Green function, when it is coupled to a field described by a continuum of modes~\cite{CohenTannoudji,Friedrichs48}. We extend the standard analysis to the case when the mode describing the ``excited state'' has an opposite norm to that of the continuum. We shall see that the analytic properties of Green functions precisely describe the consequences of this coupling. We shall then be able to obtain, this time in analytical terms, the rather complicated trajectories observed when studying the eigen-frequencies of a charged field in the electric case. 

To complete our review of the various possibilities, we also study the evolution of complex eigen-frequencies on a finite torus, when the field obeys periodic boundary conditions. In this case, QNM no longer exist, as the notion of outgoing boundary condition is no longer meaningful. Yet, many aspects previously observed when working on an infinite spatial domain are recovered. In addition, the discreteness of the spectrum allows for a simpler description of the appearance of DIM. We shall see that a pair of modes with complex frequencies, containing a DIM and its decaying partner, arises through the coalescence of two real frequency modes of opposite norms. The description of this two-mode mixing is rather similar to that leading to an avoided crossing when the two modes have the same norm.

This paper is organized as follows. In Section~\ref{sec:KG}, we present a general method for solving the Klein-Gordon equation with an electrostatic potential and apply it to study the structure of DIM in relation with the other modes. In Section~\ref{sec:Model}, we show how the model of~\cite{Friedrichs48} can be extended to describe the birth of new DIM when varying external parameters. In Section~\ref{sec:discrete}, we show the differences induced by periodic boundary conditions. We conclude in Section~\ref{Concl}. In  appendix~\ref{sec:degvsnondeg} we underline the key role played by the symplectic structure in constraining the properties of DIM. From very general considerations we show that DIM form either mode doublets or mode quartets. Both of these have been observed when studying the black hole laser effect. In appendix~\ref{App:KGmath}, we explicitly give the resolvent of the Klein-Gordon equation in 1+1 dimensions and derive a necessary condition for DIM to arise, as well as an upper bound on their growth rate, for an arbitrary electrostatic potential. Finally, in appendix~\ref{BH_App} we briefly comment on the links with instabilities in Kerr black holes.

\section{Instabilities in an inhomogeneous electric field} 
\label{sec:KG}

The principle aim of this section is to present the various types of solutions of the Klein-Gordon equation in an electrostatic potential. We shall first recover the standard continuous spectrum of real frequency modes. These modes are asymptotically bounded, i.e., their modulus is bounded because they become linear superposition of plane waves with real wave vectors at spatial infinity. Besides these, we shall encounter three other kinds of solutions which behave differently at spatial infinity, namely, \emph{quasi-normal modes} (QNM), \emph{dynamical instability modes} (DIM), and \emph{bound state modes} (BSM). 

QNM obey outgoing boundary conditions, decay exponentially in time, and are unbounded at spatial infinity. As we shall see, because of this last property they are not in the spectrum. DIM instead grow in time and are spatially bounded. In addition, we shall see that each DIM is associated with a partner mode, whose frequency is the complex conjugated, and which therefore decreases in time. These two solutions are square integrable and thus belong to the discrete spectrum of modes. The third kind of modes, BSM, possess a real frequency and are stable. They are also square integrable and thus belong to the discrete sector of the spectrum. 

As a warming up exercise we first study a simple example. It will allow us to introduce the various types of modes and the phenomenon of mode amplification. It will also allow us to describe in qualitative terms how the complex frequencies of the stationary eigen-modes evolve when varying some parameter, such as the strength of the electric field or the length of the cavity. In the second and the third parts of this Section, we expose the general theory in more mathematical terms. 

\subsection{The square potential}
\label{sub:squarepotential}

The two-dimensional model we consider is similar to that studied by Fulling~\cite{Fulling}. The key difference is that we work with an infinite spatial extension, instead of periodic boundary conditions. This is a necessary condition to obtain QNM. This case is also closer to the situations encountered when studying the stability of black holes~\cite{Kaloper}. 
Yet systems obeying periodic boundary conditions are interesting and display different features. These are studied in Section~\ref{sec:discrete} and Appendix~\ref{sec:degvsnondeg}. 

We consider the modes of a scalar complex field $\phi$ obeying the Klein-Gordon equation 
\be \label{KGE}
\lp \pd_t + i e A_0(t,x) \rp^2 \phi - \pd_x^2 \phi +m^2 \phi = 0, 
\ee
where $e$ is its charge and $m$ its mass. If $\phi_1$ and $\phi_2$ are two solutions of \eq{KGE}, the standard Klein-Gordon scalar product $\lp \phi_1,\phi_2 \rp$ is 
\be \label{Ksp} 
\lp \phi_1,\phi_2 \rp \equiv i \int \lp \phi_1^* \pi_2 - \pi_1^* \phi_2 \rp \, dx ,
\ee
where 
\be 
\pi_j \equiv \lp \pd_t + i e A_0  \rp \phi_j, \; j \in \left\lbrace 1,2 \right\rbrace \, ,
\label{scalarp}
\ee
is the conjugated momentum of $\phi_j$. The associated norm $\lp \phi_1,\phi_1 \rp$ is \emph{not} positive definite. Hence, \eqref{Ksp} does not define a positive definite scalar product, but a Krein product~\cite{Langer82,Bognar}. Importantly \eq{Ksp} is identically conserved for any pair $\phi_1,\phi_2$ of solutions of \eq{KGE}. 
Therefore, when following a particular wave packet while varying the external field $A_0$ in space or time, its norm stays unchanged. In addition, when $A_0$ only depends on $x$, there is another conserved (quadratic) quantity, given by the energy 
\be \label{Eelec} 
E[\phi] = \int \left( |\pd_t \phi|^2 + |\pd_x \phi|^2 + (m^2 - e^2A_0(x)^2)|\phi|^2 \right) dx.
\ee
In vanishing or weak potentials, one can verify that $E[\phi]$ is positive. Instead, when the potential $A_0$ is strong enough, $E[\phi]$ can become negative for some field configurations. 
This characterizes unstable systems and gives rise to the Klein paradox~\cite{Manogue88}. 
The energy is directly related to the Klein-Gordon scalar product via the relation $E[\phi] = (\phi | i\pd_t \phi)$, see Appendix~\ref{sec:degvsnondeg}. 

From now on, $A_0$ only depends on $x$. We thus look for stationary solutions of the form $e^{-i \la t} \phi_\la(x)$. These obey the stationary equation 
\be \label{eq:KGEs}
\left[ \lp \la - e A_0(x) \rp^2 + \pd_x^2 - m^2 \right] \phi_\la = 0, 
\ee
where $\la$ is a priori any complex number. Our main endeavour is to determine the mode spectrum, i.e., the set of solutions of \eq{eq:KGEs} which obey appropriate boundary conditions. As explained in the next subsection, the modes must be \emph{asymptotically bounded}. 

To simplify the discussion we shall assume that the static potential vanishes at both infinities $A_0(x) \to 0$ when $x \to \pm \infty$, but the analysis is easily generalized to different limits~\cite{Bachelot04}. To be able to obtain explicit solutions, we shall work in the simple case of a square-well potential 
\be 
A_0(x)= \left\lbrace
\begin{array}{ll}
0, & |x| > L, \\
A, & |x| < L ,
\end{array}
\right. 
\ee 
with $eA > 0$. In the three regions $I_1: x < -L$, $I_2: -L< x < L$, and $I_3: L < x$, the solutions of \eq{eq:KGEs} are plane waves $e^{\pm i k x}$. In $I_1$ and $I_3$, they have the same momentum $k_\la$, while in the internal region $I_2$, the momentum $k_\la^{\rm int}$ is changed due to the electric potential. For $\la \in \mathbb C$, one has
\bsub \be 
k_\la \eg \sqrt{\la^2 - m^2} , \label{klam} \\
k_\la^{\rm int} \eg \sqrt{(\la -e A)^2 - m^2}. \label{kintlam} 
\ee \esub
To obtain globally defined solutions of \eq{eq:KGEs}, we must impose continuity of $\phi$ and $\pd_x \phi$ at $x = \pm L$.

\subsubsection{The mode spectrum in a single-step potential} 

Before discussing the spectrum for the square-well potential, we first consider the simpler case of a single step: 
\be \label{step} 
A_0(x) = \theta(-x) A, 
\ee
where $\theta(x)$ is the Heaviside function. 
As we shall see, this setup exhibits the main ingredients which will generate the DIM in a square potential. \eq{step} implies that the wave-vectors of the stationary modes of frequency $\la$ are $\pm k_\la$ of \eq{klam} on the right of the jump, and $\pm k_\la^{\rm int}$ of \eq{kintlam} on its left. To characterize the outgoing modes, we choose the sign of the square roots so that $\Im k_\la \, \Im \la \geq 0$, and  $\Im k_\la^{\rm int} \, \Im \la \geq 0$. Then, the branch cut of $k_\la$ is the segment $[-m;m]$, and that of $k_\la^{\rm int}$ is $[- m+ eA;m+eA]$. Hence, since $\pd_\la k_\la > 0$ for $\la \in \mathbb{R} \setminus  [-m,m]$, solutions of \eq{eq:KGEs} obey the {\it outgoing boundary condition}, i.e., their group velocity is oriented away from the potential 
step, if they are proportional to $e^{+ i k_\la x}$ for $x > 0$.  Similarly, for $x < 0$, the wave $e^{-i k_\la^{\rm int} x}$ is outgoing when $\la \in \mathbb{R} \setminus  [-m+ eA,m+eA]$. 

These observations imply that the mode which obeys outgoing boundary conditions on the left side has the following form 
\be 
\phi_\la(x) = e^{-i k_\la x} + R_\la \, e^{i k_\la x} & x>0 \nonumber \\
\phi_\la(x) = T_\la \, e^{ - i k_\la^{\rm int}  x} & x<0, 
\label{scat}
\ee 
where  $R_\la$ and $T_\la$ respectively give the reflection and the transmission coefficients. When $\la \in \mathbb{R} \setminus \lp [-m,m] \cup [e A -m, e A +m] \rp$, conservation of the Wronskian 
$W = \phi_\la^*  \partial_x \phi_\la -\phi_\la  \partial_x \phi_\la^* $ implies that 
\be
|R_\la|^2 + |T_\la|^2 \, \Re(k_\la^{\rm int} / k_\la) = 1 \, . 
\label{Wr}
\ee
This mode is asymptotically bounded when $\la \in \mathbb{R} \setminus  [-m,m]$. Then, $k_\la$ is real, so the mode for $x>0$ is a superposition of two plane waves. For $x<0$, the transmitted mode is also a plane wave if $\la \notin (e A - m, e A + m)$, while it is exponentially decreasing for $\la \in (e A - m, e A + m)$, as $k_\la^{\rm int} = i \sqrt{m^2 - (\la - e A)^2}$. For $\la \in \mathbb{C} - \mathbb{R}$, the mode of \eq{scat}  is in general not asymptotically bounded. Indeed, as soon as $\Im \la \neq 0$, $k_\la$ and $k_\la^{\rm int}$ acquire a non-vanishing imaginary part, so that the mode will grow exponentially at infinity. The mode obeying outgoing boundary conditions on the right region is obtained from \eq{scat} by the transformation $x \to -x, k_\la \leftrightarrow k_\la^{\rm int}$. It is asymptotically bounded when $\la \in \mathbb{R} \setminus  [e A-m,e A +m]$. 

\begin{figure} 
\includegraphics[width=0.49 \linewidth]{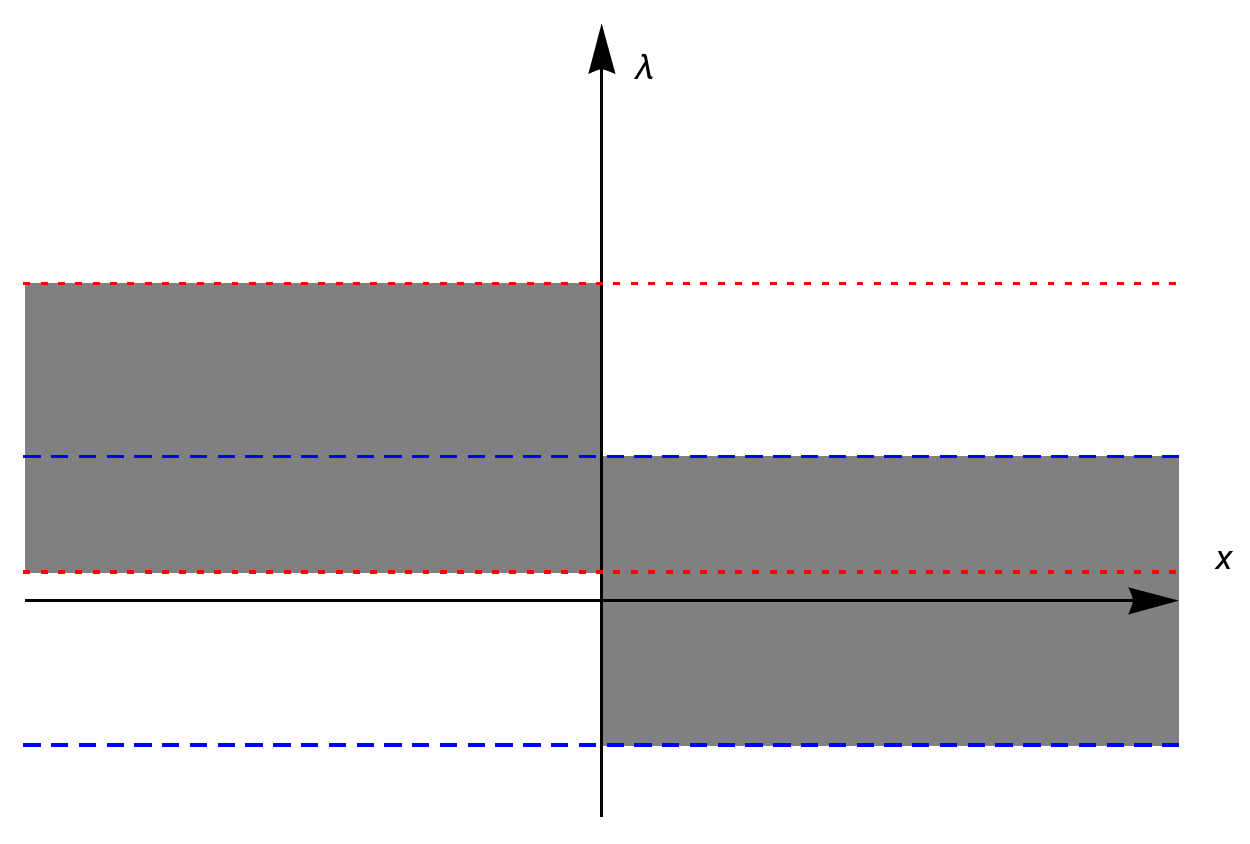}
\includegraphics[width=0.49 \linewidth]{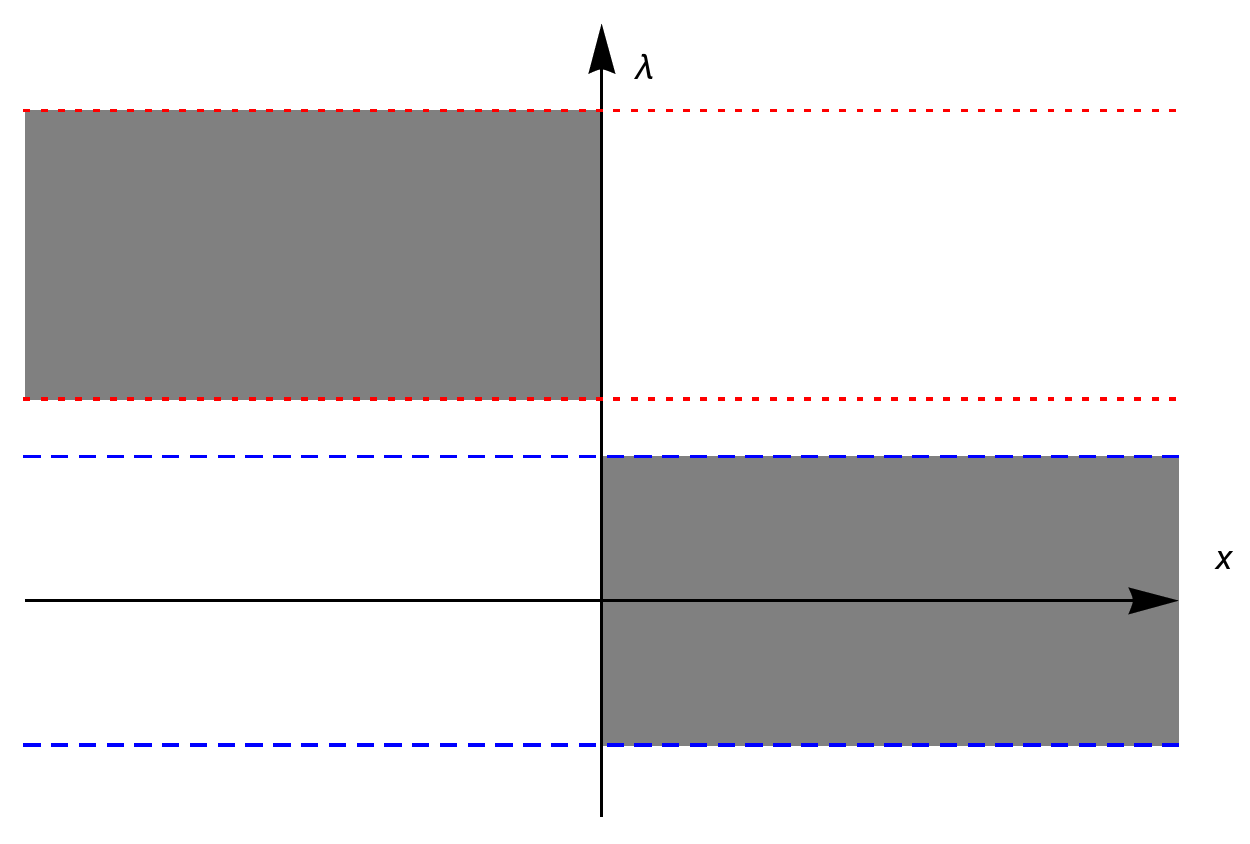}
\caption{We show the forbidden domains of $\la$ as functions of $x$ in the case of a step-like electrostatic potential localized in the left region. The gray areas correspond to the forbidden bands given the mass $m$. The two dashed, blue lines show $\la = \pm m$ and the two dotted, red ones show $\la = e A \pm m$. On the left plot, $e A = 1.2 m$, so there is no Klein region. On the right plot, $e A = 2.4 m$ and a Klein region is present between the upper blue and lower red lines.} \label{eAvs2m} 
\end{figure}

A deeper understanding of these mathematical properties is obtained when considering the sign of the norm of \eq{Ksp}. Two very different types of situations can be found. The standard case is a smooth deformation of the translation invariant case where the potential vanishes. It is found when the potential barrier is sufficiently low: $0 \leq e A < 2 m$, see the left plot of Fig.~\ref{eAvs2m}.
In this case, the norms of the plane waves $e^{\pm i k_\la x}$, $e^{i k_\la^{\rm int} x}$ have the same sign. When working with positive frequencies $\la > e A + m$ (above the dotted red line in the left plot of Fig.~\ref{eAvs2m}), they are positive and \eq{scat} describes the elastic scattering of a positively charged particle. The positive quantities $|R_\la|^2$ and $|T_\la|^2 (k_\la^{\rm int} / k_\la)$ are respectively interpreted as the probability for an incoming particle from the right to be reflected or transmitted through the electro-static barrier. When $m < \la < e A + m$ one finds that there is a total reflection to the right, i.e., $|R_\la|=1$, as the particle cannot propagate (on-shell) in the left region, as described above.  

In the interval $\la \in [-m + eA,m]$, there is no asymptotically bounded mode: any solution grows exponentially on one or the other side. Below this forbidden band, for $\lambda <  - m + eA < m$, there are again asymptotically bounded modes satisfying outgoing boundary conditions (at least on one side). However, the norm of the plane waves it contains is now negative. In second quantized settings these negative norm modes are associated with creation operators of anti-particles (i.e., negatively charged particles)~\cite{Manogue88}. For $\lambda < - m$, the mode of \eq{scat} then describes the scattering of an anti-particle coming from the right. Notice that its energy has the {\it opposite} sign of $\lambda$. (In units where $\hbar = c = 1$ the energy of this field excitation would be is given by $- \lambda > m $). This sign flip between frequency and energy found for negative norm modes is essential, and also applies to classical fields. In brief, combining positive and negative norms, the frequency spectrum of asymptotically bounded modes is real, continuous, and given by $\la \in \mathbb{R} \setminus [-m+ eA,m]$. 

The unusual case (giving rise to the ``Klein paradox'') is found when there is a Klein region, i.e., when the jump of $eA$ is larger than $2m$. In this case, there exists a range of frequencies, clearly visible in the right plot in Fig.~\ref{eAvs2m},
\be
m < \lambda < e A -m ,
\label{Kleinrange}
\ee 
where the transmitted wave of \eq{scat} has a negative norm, although it emerges from an incoming positive norm mode propagating in the right region. In fact, with our convention for the sign of the square root, the ratio $(k_\la^{\rm int} / k_\la)$ of \eq{Wr} is negative. Hence \eq{Wr} implies that $|R_\la|^2$ is larger than $1$. There is no paradox here as conservation of the Wronskian $W$ 
must be interpreted as describing the charge conservation associated with the creation of a negatively charged particle emitted to the left~\cite{Manogue88}.~\footnote{When dealing with a neutral field, one  looses this interpretation based on the charge. However, one keeps the fact that any mode amplification occurring in a stationary situation (i.e., in the presence of a stationary Killing field) involves waves carrying equal and opposite (Killing) energy (unless the real part of the frequency vanishes, as it is the case for the black string~\cite{Gregory93}, see the discussion of Appendix~\ref{sec:degvsnondeg}.) Hence, charge conservation is only a useful guide, and not an essential ingredient. Instead, the conservation of the scalar product and that of the hamiltonian (the field energy), both non-positive definite, are essential. Because the norm of real (classical) waves of hermitian fields identically vanishes, it is the sign of the wave energy which should be used to identify the modes involved in dynamical instabilities and super-radiance.\label{waveen}} 

The scattering coefficients relating incoming modes to outgoing ones in \eq{scat} clearly distinguish the two cases, namely the elastic scattering, $|R_\la|^2 < 1$, and the mode amplification, $|R_\la|^2 > 1$. Their values are fixed by the matching conditions at $x=0$, i.e., continuity of $\phi$ and $\pd_x \phi$. In the present case, one gets 
\be \label{SuperradCoef}
R_\la = \frac{k_\la - k_\la^{int}}{k_\la + k_\la^{int}}, \; T_\la = \frac{2 k_\la}{k_\la + k_\la^{int}}.
\ee
When $\la$ is outside the interval of \eq{Kleinrange}, there is either a total or a partial reflection. In the second case, $k_\la$ and $k_\la^{int}$ 
are real and have the same sign so $|R_\la|^2$ and $|T_\la|^2 (k_\la^{\rm int} / k_\la)$ are both smaller than $1$, and sum up to 1.
(When there is total reflection, one of these two wave-vectors is purely imaginary.) The $S$-matrix relating incoming to scattered modes is thus given by an element of $U(2)$. Instead when $\la \in (m, e A -m)$, the two wave-vectors, and their contributions to the Wronskian of \eq{Wr}, have opposite signs. Thus $|R_\la|^2 $ is necessarily greater than 1 (while $|T_\la^2 (k_\la^{\rm int} / k_\la)|$ may be either smaller or larger than $1$). In this case, the $S$-matrix is an element of $U(1,1)$. 

In brief, when there is a Klein region, the spectrum of stationary modes contains a {\it continuous} but finite frequency range given by \eq{Kleinrange} wherein modes with negative norm (describing negative charged particles with energy smaller than $-m$) mix with the standard continuum of modes with positive norm, charge, and energy larger than $m$. As a result, in this frequency range there is a steady pair creation rate which preserves the total energy (and the total charge).

\subsubsection{The mode spectrum in a square well}
\label{subsub:modessquarewell}

When considering a square well of finite extension the spectrum of modes is radically different. Indeed, the continuous spectrum of positive and negative norm modes co-existing in the frequency interval of \eq{Kleinrange} for a single barrier no longer exists. As we shall see, it is replaced by a discrete set of DIM which have a vanishing norm. This discrete set is obtained by requiring that the modes be square integrable. Interestingly, it can also be obtained by imposing purely outgoing boundary conditions~\cite{Kokotas99,Berti09}. Then, when $\Im \la > 0$, the mode grows in time and decays in space. Hence it is a DIM. Instead, when $\Im\la < 0$ it decays in time but is not asymptotically bounded. Hence it is a QNM. A more rigorous justification of these statements will be given in the next subsections. 

To get the discrete spectrum of eigen-frequencies $\la$, we again choose the branch cut of $k_\la$ on the segment $[-m;m]$. Under this convention, as explained above, $\pd_\la k_\la > 0$ for $\la \in \mathbb{R} \backslash [-m,m]$ so that solutions of \eq{eq:KGEs} obey outgoing boundary conditions if they are proportional to $e^{-i k_\la x}$ in $I_1$ and to $e^{i k_\la x}$ in $I_3$. Imposing continuity of $\phi$ and $\pd_x \phi$ at $x = \pm L$, we find that such a solution exists under the condition 
\be \label{efe} 
\left(\frac{k_\la}{k_\la^{\rm int}}-1\right)^2e^{2i k_\la^{\rm int} L} = \left(\frac{k_\la}{k_\la^{\rm int}}+1\right)^2e^{-2i k_\la^{\rm int} L}.
\ee

\begin{figure}[h!] 
\begin{center}
\includegraphics[width=0.45\textwidth]{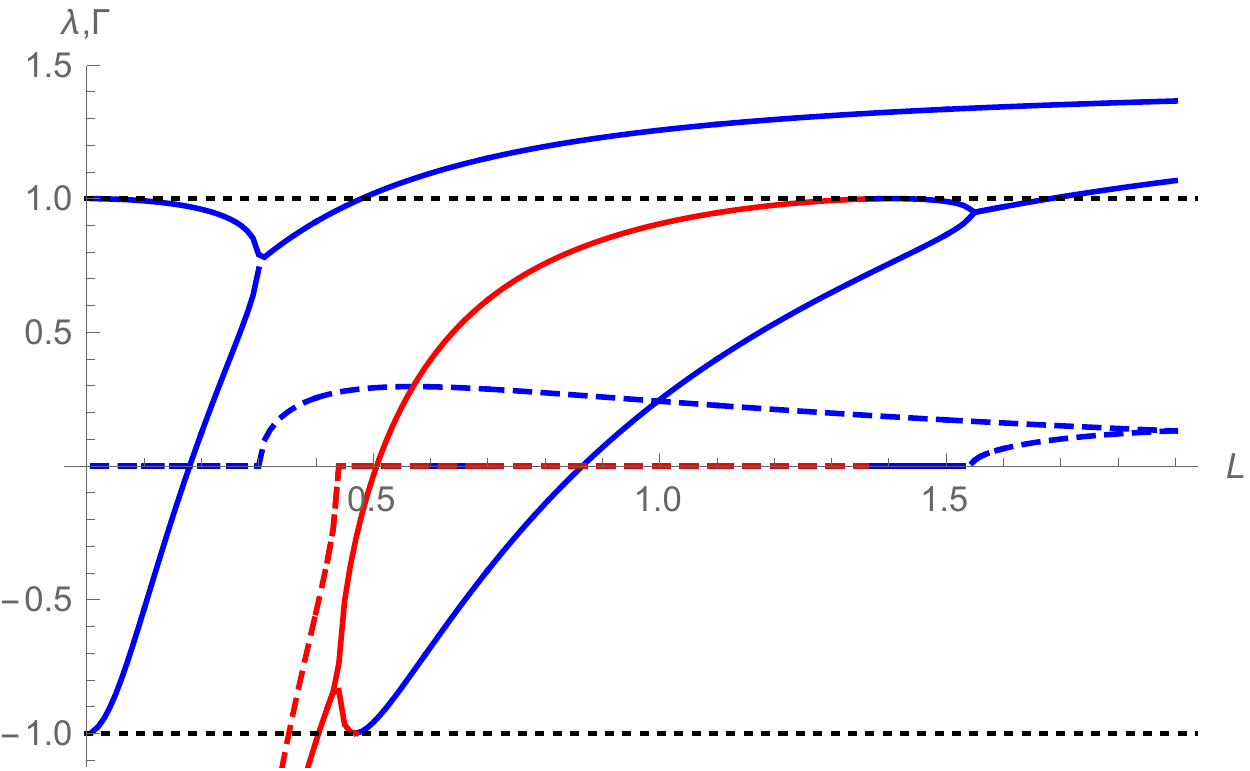} 
\includegraphics[width=0.45\textwidth]{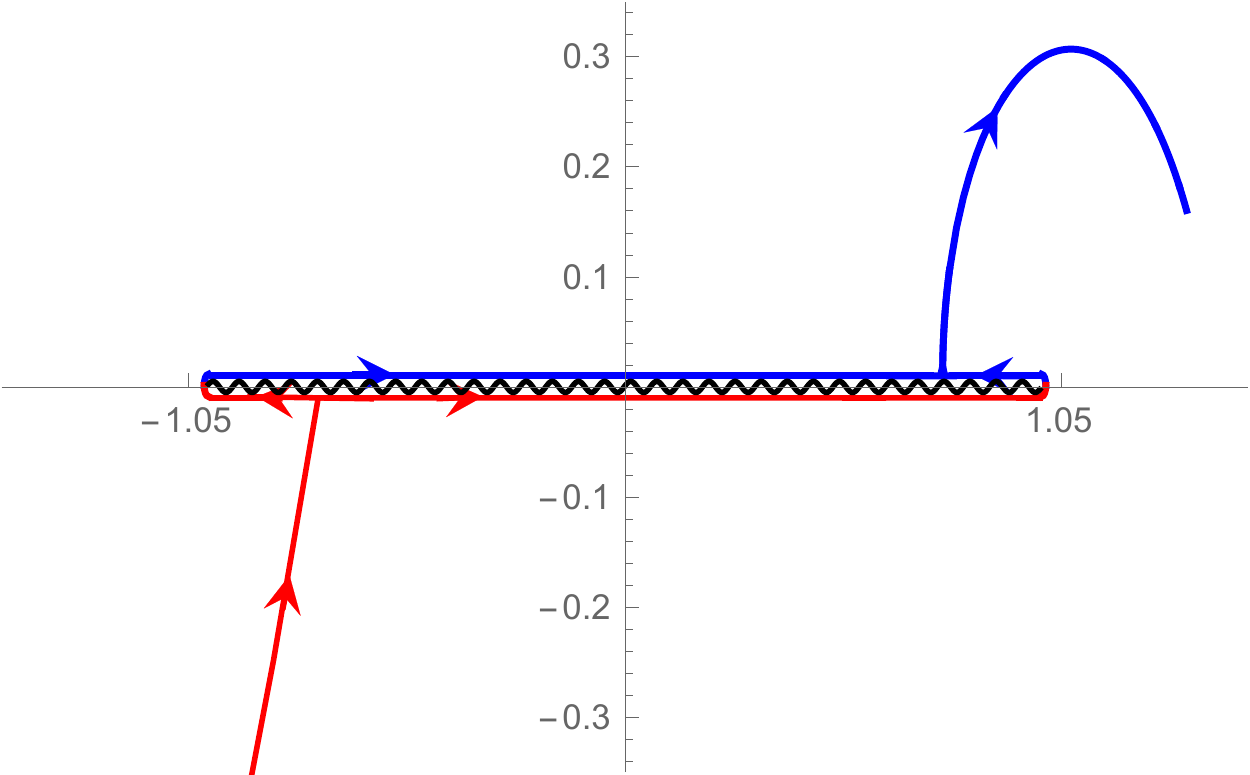}
\caption{Plot of the frequencies the first two DIM in the massive case, with $m=1$ and $A=2.5$. On the left panel, we represent their real (plain line) and imaginary (dashed line) parts. One clearly sees that the first DIM appears when $L$ is slightly larger than $0.3$ while the second DIM appears for $L$ larger than $\sim 1.53$. The two dotted lines represent the mass gap (i.e., the lines $\om = \pm m$). On the right panel, we represent the trajectory of the second frequency in the complex plane as $L$ increases. The lines are red when the root lies in the lower half-plane ($\Im(\la) \leqslant 0$) and blue in the upper half-plane ($\Im(\la) \geqslant 0$). Hence the modes described by the roots in red are QNM, whereas those described by the roots in blue are BSM when $\Im \la = 0^+$ and DIM when $\Im \la > 0$. 
} \label{fig:Roots_Elec_Massive} 
\end{center}
\end{figure}

For a massive field, there is no DIM when $|e A| < 2 m$. Indeed, as indicated in \eq{Kleinrange}, the non-trivial mode mixing arises only if the electric field is sufficiently strong for the energy of the anti-particle to drop below $-m$, see also appendix~\ref{app:Klein}. In addition, as shown in Fig.~\ref{fig:Roots_Elec_Massive}, $L$ must be larger than a certain threshold value for having DIM in the spectrum. 

The first DIM appears when two real frequency modes merge. Before this merging, the solution with a frequency slightly below $m$ is a positive norm mode which describes a positively charged particle in a bound state modes (BSM). Instead the solution with an initially negative frequency equal to $-m$ is a negative norm mode which describes a negatively charged particle in a BSM. When their frequencies merge, the energy cost to create a pair of these excitations vanishes. For greater values of $L$, the two BSM are replaced by a pair of eigen-modes with complex conjugated frequencies. It can be shown that each of them has a vanishing norm, and a vanishing contribution to the energy. Yet they have a non-vanishing overlap with each other. For more details about this, we refer to~\cite{Coutant10,Manogue88}. When keeping increasing $L$, this pair of modes persists. (When considering the square well on a torus, one finds that DIM can disappear for certain values of $L$, see Section~\ref{sec:discrete} and \cite{Fulling}.)

When $L$ grows, we also see that a new DIM appears. Interestingly, it appears through a different scenario than the first. When a complex frequency QNM (in red) reaches the real axis, it splits into two modes with real frequencies $\la_1, \la_2 \in [-m,+m]$.~\footnote{\label{BSMvsQNM} Because of the discontinuity in $k_\la$ across the segment $[-m,m]$, one should clearly distinguish if $\la$ lies above or below the branch cut. Indeed, close to the real axis, for $\Re \la \in (-m,m)$, the imaginary part of $k_\la$ is given by $\Im(k_\la) = \text{sign}(\Im(\la)) \sqrt{m^2 - (\Re\la)^2}$. Hence, outgoing modes associated with a frequency above the branch cut decay asymptotically, and are thus BSM. Instead, those associated with a frequency below the cut asymptotically diverge and are QNM.} A close examination of \eq{efe} shows that they arrive below the branch cut, then travel along the branch cut in opposite directions, pass onto the upper side, and finally merge there (see right panel of Fig.~\ref{fig:Roots_Elec_Massive}). When they merge, they give birth to a DIM. 
Further numerical simulations (not shown in \fig{fig:Roots_Elec_Massive}) show that other DIM periodically appear when keeping increasing $L$. These additional DIM all appear through the same process as the second one.

To understand what happens in analytical terms, it is instructive to consider the ``deep well limit'', i.e. $eA \to +\infty$ at fixed $L$ and $m$. In this limit, the DIM and QNM frequencies are given by the following Taylor expansion in $1/eA \ll 1$: 
\be 
\la_n = e A \mp \sqrt{m^2+\frac{\pi^2 n^2}{4 L^2}} \pm i \frac{\pi^2 n^2}{2 L^2 e A \sqrt{4 m^2 L^2+\pi^2 n^2}}+\mathcal{O} \lp \frac{1}{(e A)^2} \rp,
\ee
where the upper signs correspond to DIM and the lower signs to QNM, and $n$ is an integer. The condition $\la_n = 0$ approximately gives the dimensionality $N$ of the set of DIM. One finds here $N \sim 2LeA/\pi$. They correspond to the eigen-modes of the cavity between $x = \pm L$. The imaginary part is given by the third term. It arises from the coupling of the cavity mode with the exterior modes. In fact, it is exactly given by the semiclassical result $\Delta t_{\rm rt} \Im \la_n = |\beta|^2$
(see Eq.~(41) in \cite{Coutant10}, with $\cos \psi_a$ = 0 in the present case), where $\Delta t_{\rm rt}$ is the semiclassical time for an anti-particle to make a round trip in the cavity ($\sim 4L/v_g$) and $|\beta|^2$ the production rate at each barrier~\cite{Manogue88}. It is given by $|T_\la|^2 (k_\la^{\rm int} / k_\la)$, where $T_\la$ is the transmission coefficient of of \eq{SuperradCoef}. 

\begin{figure}[h!] 
\begin{center}
\includegraphics[width=0.45\textwidth]{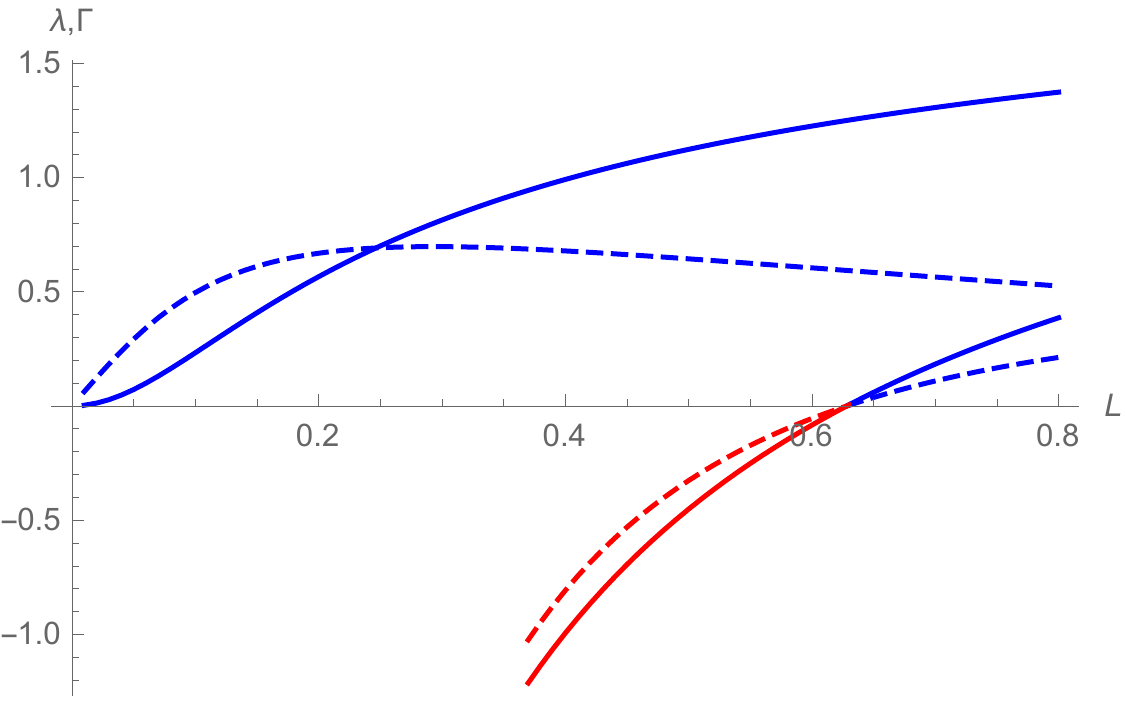}
\includegraphics[width=0.45\textwidth]{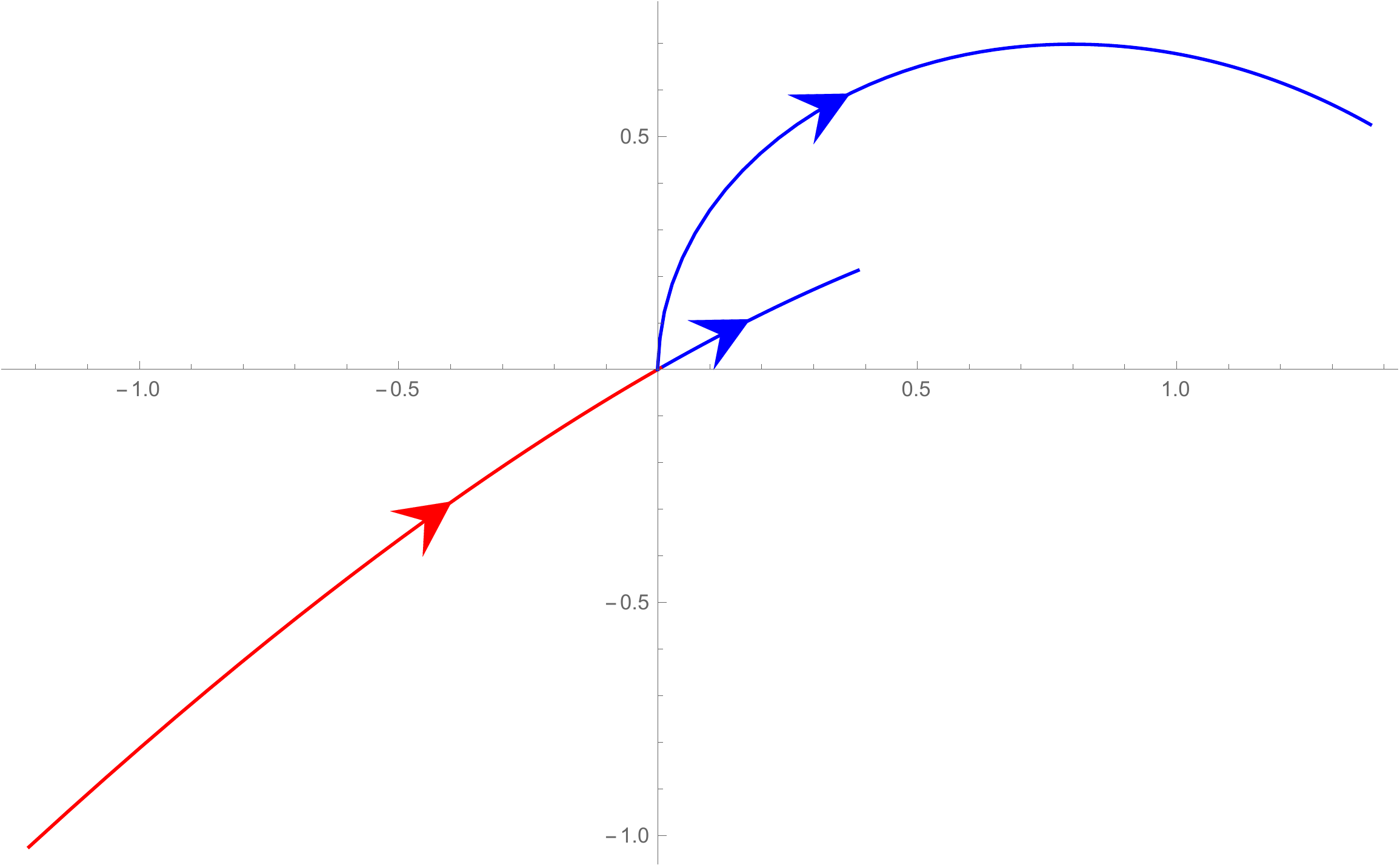}
\caption{Plot of the two first complex frequencies $\la$ associated with the outgoing solutions of \eq{efe} for a massless field with $e A=2.5$. On the left panel, we represent their real (plain line) and their imaginary (dashed line) parts. On the right panel, we represent the trajectory of these two eigen-frequencies in the complex plane. Arrows indicate the direction of increasing $L$, from $0$ till $0.8$. One clearly sees that the second DIM arises from a QNM (in red).
} \label{fig:Roots_Elec_Massless} 
\end{center}
\end{figure}

It is also instructive to consider the massless case, which shows different features. The propagation of the first eigen-frequencies is given in Fig.~\ref{fig:Roots_Elec_Massless}. 
As soon as $eA \neq 0$, there is a DIM for all values of the extension $2L > 0$. This is due to the fact that the potential can host a bound state of an anti-particle with a negative energy. When considering modes with $\Re \lambda > 0$, it is described by a negative norm mode. As in the case of a single barrier, it will mix with the continuum of positive norm modes describing positively charged particles reaching infinity. Because the negative energy mode is trapped, the higher its amplitude, the faster is the pair production, resulting in an emission that exponentially grows in time. 

In addition, when increasing $L$, the half-length of the trapped region, the energy of the bound states diminishes. When it is still positive, the associated mode mixes with the continuum spectrum of modes describing negative charge positive energy anti-particles. In this case, as we shall see in Section~\ref{sec:Model}, the coupling gives rise to a QNM. This means that the bound state will decay into some anti-particle state. This is the situation represented in red which exists for $L< 0.6$ in Fig.~\ref{fig:Roots_Elec_Massless}. However, each time $L$ crosses the critical values  
\be 
L_{n} = \frac{\pi n}{2 e A}, \qquad (n \in \mathbb N), 
\ee 
the energy of the $n^{\rm th}$ bound state becomes negative (i.e., the frequency of the negative norm mode becomes positive). As a result, the negative norm bound mode now couples to the continuum of positive norm modes. Hence the QNM has evolved into a DIM. 

\begin{figure}
\begin{center}
\includegraphics[width=0.49 \linewidth]{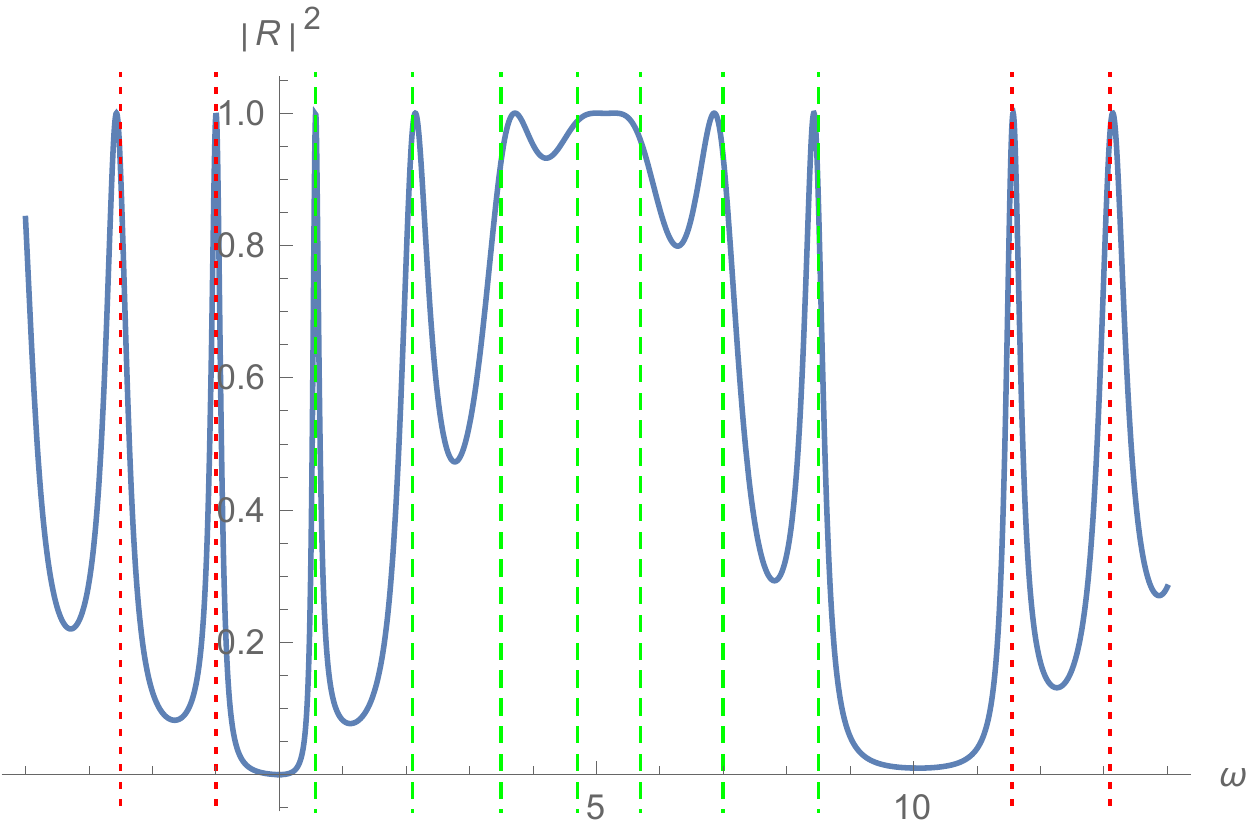}
\includegraphics[width=0.49 \linewidth]{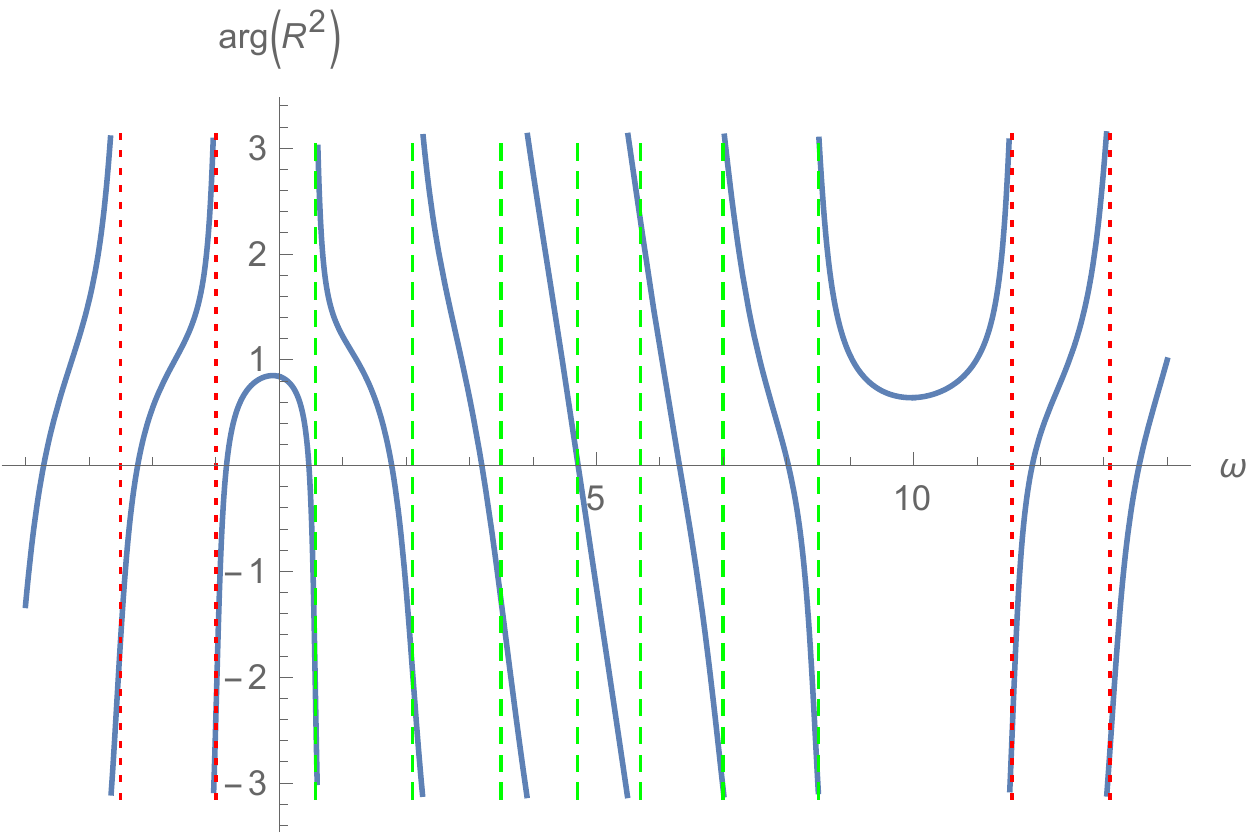}
\end{center}
\caption{Plots of the modulus (left) and the phase (right) of the squared reflection coefficient $R_\omega^2$, for real frequencies $\omega$. Dotted, red (dashed, green) lines show the real parts of the QNM (DIM) frequencies. The electrostatic potential is equal to $e A = 10$ in a region of length $2L = 2$.}\label{fig:R}
\end{figure}

As shown in~\cite{Finazzi10} in the context of black-hole lasers in Bose-Einstein condensates, DIM reveal themselves as peaks of the overlap between scattering modes with real wave vectors and the negative norm mode living in the inside region (which plays the role of the Klein region in our settings). They also manifest themselves as a rapid change in the phase of the reflection coefficient. Interestingly, the phase shifts allow to distinguish DIM and QNM. Indeed, the phase  \emph{increases} by $\pi$ for a QNM, whereas it \emph{decreases} by the same amount for a DIM~\cite{Zapata11}. We verified that these results hold in the present setup, see Fig.~\ref{fig:R}. The continuous curves of this figure show the norm (left panel) and the argument (right panel) of $R_\om^2$, the squared reflection coefficient for an incoming mode from the left. This coefficient can be computed by simply matching inside and outside modes, and imposing that the coefficient of the left-moving wave on the right vanishes. We obtain
\be 
R_\om = 2 e^{-2 i k_\om L} \lp 2 \cos \lp 2 k_\om^{\rm int} \rp - i \lp \frac{k_\om^{\rm int}}{k_\om} - \frac{k_\om}{k_\om^{\rm int}} \rp \sin \lp 2 k_\om^{\rm int} \rp \rp^{-1}.
\ee
On the left panel of Fig.~\ref{fig:R}, 
the width of the peaks  accurately gives the imaginary part of the QNM and DIM frequencies. 
The increases (QNM) and decreases (DIM) of the argument of $R_\omega^2$ when increasing $\omega$ are clearly seen on the right plot. 

\subsection{Eigen-mode decomposition and the status of complex frequency modes}
\label{sub:modebasis}

We now present a general resolution of the Klein-Gordon equation in terms of frequency eigen-modes. The mode spectrum will be identified from the singularities of the stationary Green function obeying certain boundary conditions.

In the presence of a Krein product rather than a positive definite scalar product, one cannot appeal to a general spectral theorem that would imply that the eigen-values $\lambda$ of \eq{eq:KGEs} are all real~\cite{ReedSimon,LevyBruhl}. Yet, in the case of the 1+1 dimensional Klein-Gordon equation, there exists an explicit construction~\cite{Bachelot04}. Following that work, we consider a solution $\phi(x,t)$ of \eq{KGE}, and define its Laplace transform
\be 
\tilde{\phi}(x; \la) \equiv \int_0^\infty \phi(x,t) \, e^{i \la t} \, dt,
\ee 
for $\la$ complex such that $\Im \la > 0 $ is sufficient large so that the above integral converges. From this definition, $\tilde{\phi}(x,\la)$ obeys the inhomogeneous equation 
\be \label{eq:Laplacephi} 
\left[ (\la - e A_0 (x))^2 + \pd_x^2 - m^2 \right] \tilde{\phi}(x,\la) = -\pi(x,0)+i (\la - e A_0(x)) \phi(x,0) \equiv J_\la^0(x). 
\ee
Therefore, we now need to solve the stationary mode equation with a source term given by the initial conditions. The time-dependent solution of \eq{KGE} is then obtained using the inverse Laplace transform
\be \label{positive_t}
\Theta(t) \, \phi(x,t) = -\frac{1}{2 \pi} \int_{D_{\rm ret}} \tilde{\phi}(x; \la) 
\, e^{-i \la t} \, d \la,
\ee
where $D_{\rm ret}$ is a contour in the complex $\la$-plane above all singularities of $\tilde{\phi}(x; \la)$. To get the field for all times, and not only $t>0$, we define $\tilde{\phi}(x; \la)$ for $\Im \la $ sufficiently negative as the advanced Laplace transform. The corresponding inverse is obtained with the help of the advanced contour $D_{\rm adv}$, lying below all singularities. Combining this with \eq{positive_t}, we get
\be \label{eq:contourphi}
\phi(x,t) = -\frac{1}{2 \pi} \int_{D_{\rm ret} \cup D_{\rm adv}} \tilde{\phi}(x; \la) \, e^{-i \la t} \, d \la. 
\ee
This contour integral has the advantage of being symmetric between the lower and upper half complex $\la$-plane. To solve \eq{eq:Laplacephi} and obtain $\tilde{\phi}(x;\la)$, we introduce the Green function $G_\la(x,x')$. $\tilde{\phi}(x;\la)$ is then given by 
\be \label{eq:GJint}
\tilde \phi(x;\la) = \int G_\la(x,x') \, J_\la^0(x') \, dx'.
\ee
The Green function satisfies the mode equation sourced by a Dirac function, i.e., 
\be \label{eq:G}
\left[ \lp \la - e A_0(x) \rp^2 + \pd_x^2 - m^2 \right]G_\la(x,x') = \delta(x-x'). 
\ee
Because this is a second order differential equation, there exists a simple construction in terms of Jost functions~\cite{Newton}. A pair of Jost functions consists of two linearly independent solutions 
$\phi_\la^{(1)}$ and $\phi_\la^{(2)}$ of the stationary mode equation \eqref{eq:KGEs}. The Green function is then built as their space-ordered product 
\be \label{Gfunction}
G_\la (x,x') = \frac{1}{W(\la)} \left\lbrace
\begin{array}{ll}
\phi_\la^{(1)}(x) \, \phi_\la^{(2)}(x') \quad \text{if} \quad x < x' \\
\phi_\la^{(1)}(x') \, \phi_\la^{(2)}(x) \quad \text{if} \quad x' < x 
\end{array}
\right. , 
\ee 
where $W(\la)$ is the Wronskian
\be 
W(\la) = \phi_\la^{(1)} \pd_x \phi_\la^{(2)} - \phi_\la^{(2)} \pd_x \phi_\la^{(1)}.
\ee
$G_\la$ is then uniquely determined by a choice of boundary conditions, which select the Jost functions $\phi_\la^{(1)}$ and $\phi_\la^{(2)}$. Here we require that the Green function decays sufficiently fast for $x \to \pm \infty $ so that the integral of \eq{eq:GJint} always converges. This is a necessary condition to obtain localized solutions of the time-dependent equation \eqref{KGE}~\footnote{\label{lcz} By ``localized'', we mean such that $\int (|\phi|^2 + |\pi|^2)dx < \infty$. This quantity is not conserved in time, however, it defines the topology of the Krein space associated with \eqref{Ksp}~\cite{Bognar,Langer82}. (Note that this choice is not unique~\cite{Gerard12}, but it appears as the simplest to us.) The boundary conditions of \eq{eq:BC} ensure that this condition is always fulfilled. 
}. This imposes the boundary conditions 
\be \label{eq:BC} 
\left\lbrace
\begin{array}{ll}
\phi_\la^{(1)} \; \text{is} \; L^2 \; \text{for} \; x \to  -\infty, \\
\phi_\la^{(2)} \; \text{is} \; L^2 \; \text{for} \; x \to +\infty.
\end{array}
\right.
\ee
These conditions are stronger than the asymptotic boundedness used in the former subsection. 

\begin{figure}[h!]
\centering
\includegraphics[width=0.9\linewidth]{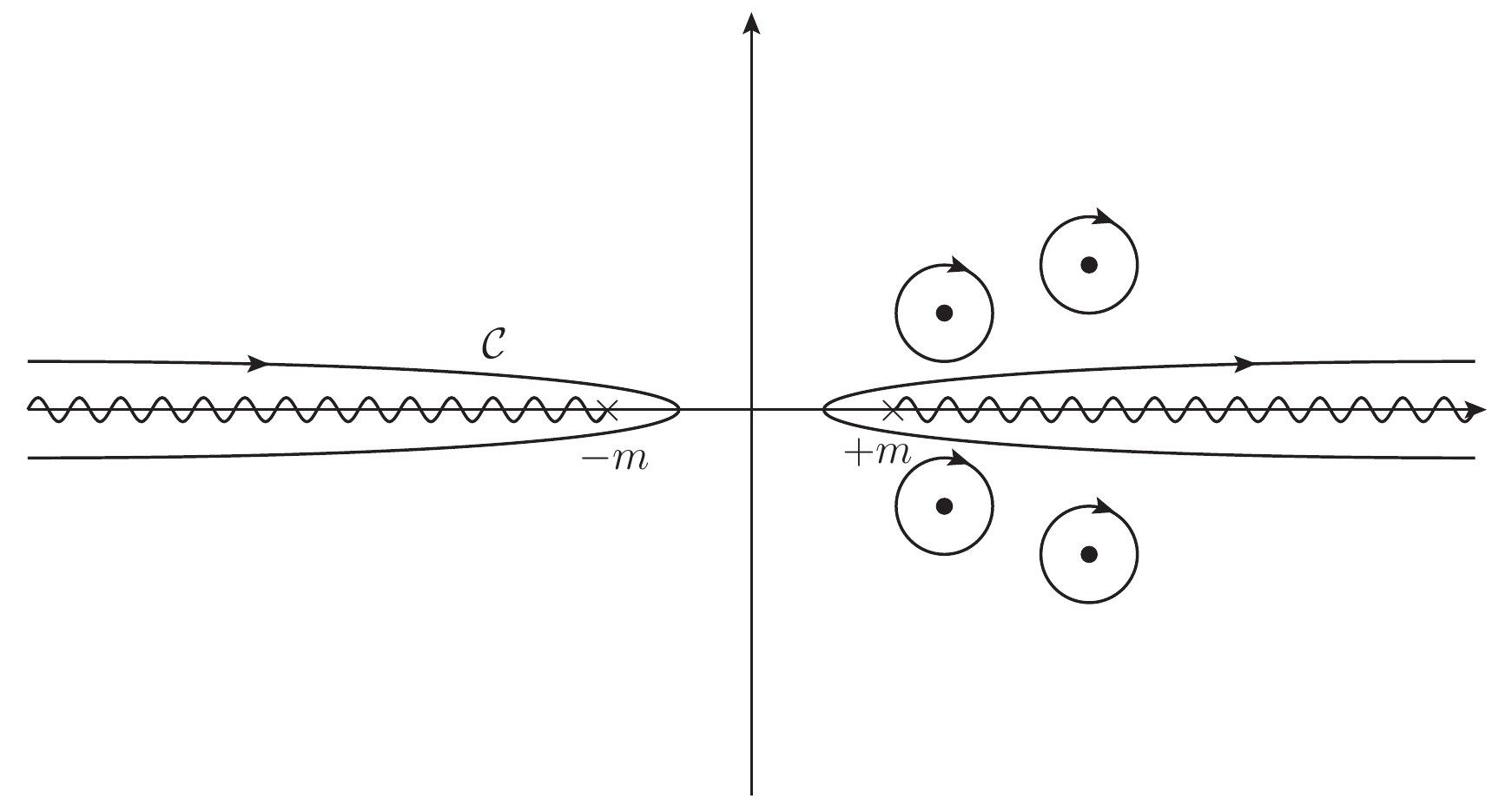}
\caption{Integration contour of \eq{eq:phit} in the physical plane. The wavy lines give the branch cuts of $G_\la$. The associated modes have a real frequency larger than $m$. They describe the scattering modes. The dots are the complex eigen-values, appearing in pairs of complex conjugates. This discrete set describes the DIM, but not the QNM. It should be noticed that the contour and the singularities are unchanged under complex conjugation $\la \to \la^*$, see \eq{Gccsym}. 
}
\label{fig:phys_contour}
\end{figure}

To obtain the time dependent solution, we combine Eqs.~\eqref{eq:contourphi}, \eqref{eq:Laplacephi}, and \eqref{eq:GJint} to get 
\be \label{eq:phit}
\phi(x,t) &=& -\frac{i}{2\pi} \int_{\mathbb R} \left( \int_{D_{\rm ret} \cup D_{\rm adv}} e^{-i\la t}(\la - e \, A_0(x'))G_\la(x,x')d\la \right) \phi_0(x) dx', \nonumber \\ 
&&+ i \int_{\mathbb R} \left( \int_{D_{\rm ret} \cup D_{\rm adv}} e^{-i\la t} G_\la(x,x')d\la \right) \pi_0(x) dx' .  
\ee
The analytic properties of $G_\la$ allow us to deform $D_{\rm ret} \cup D_{\rm adv}$ into a contour $\mathcal{C}$ encircling the spectrum, that is, singularities of $G_\la$ (see Fig.~\ref{fig:phys_contour}). Using the asymptotic conditions on the potential in \eq{eq:KGEs}, the structure of the spectrum is quite easy to characterize. In the square potential example of section \ref{sub:squarepotential}, we worked with $A_0(x)$ of compact support, hence, for $x$ large enough, any solution of \eq{eq:KGEs}  is a superposition of two plane waves: 
\be 
\phi_\la(x) \underset{x \to \pm \infty}{=} a_\pm e^{i k_\la x} + b_\pm e^{-ik_\la x}, 
\label{eq:asmode}
\ee
where $k_\la$ is defined in \eq{klam}. When $\la$ is real and $\la^2 > m^2$, $k_\la$ is real, so the modes of \eq{eq:asmode} are asymptotically bounded but not integrable, i.e., they do not satisfy \eq{eq:BC}. Conversely, asymptotically bounded modes that are not $L^2$ exist only when $\la^2 > m^2$. Therefore, the continuous spectrum is given by $\la \in \mathbb{R} \setminus ]-m,m[$, and the corresponding solutions of \eq{eq:KGEs} are scattering states. When $\la$ is not in this set, $k_\la$ possesses a non zero imaginary part. Therefore, one must impose that some of the coefficients $(a_\pm, b_\pm)$ in \eqref{eq:asmode} vanish so as to satisfy the boundary conditions \eqref{eq:BC}. In this case, the construction of $G_\la$ directly follows, unless $W(\la)$ vanishes, in which case $G_\la$ of \eq{Gfunction} has a pole. Once the Jost functions satisfying \eqref{eq:BC} have been obtained, $W$ is an analytic function of $\la$~\cite{Bachelot04}, so $W$ can only vanish for a discrete set. In this case, $\phi_\la^{(1)}$ and $\phi_\la^{(2)}$ are proportional, meaning that there exists a solution of \eq{eq:KGEs}  which decays on both sides, i.e., a solution which is $L^2$ on both sides.

If $\la$ is complex, the corresponding mode necessarily describes a dynamical instability. Its unstable character is guaranteed by the fact that complex frequency modes that are $L^2$ on both sides appear as pairs with complex conjugated frequencies. Hence one of them will grow exponentially in time. This pairing directly follows from the identity 
\be \label{Gccsym}
\left[G_\la(x,x')\right]^* = G_{\la^*}(x,x'), 
\ee
which is a consequence of the fact that the boundary conditions \eq{eq:BC} are unchanged under complex conjugation.\footnote{This property is a direct consequence of the self-adjointness of the evolution operator. In the case of the Schr\"odinger equation, with a positive definite scalar product, this condition guarantees that the spectrum is purely real. In our case, the Klein-Gordon product \eqref{Ksp} being nonpositive can vanish, and therefore the reality of $\la$ is no longer guaranteed.} This  
implies that whenever $G$ has a pole at $\la$, it has also a pole at $\la^*$. It should also be noticed that the norm of DIM necessarily vanishes as the conservation of the scalar product implies
\be
(\la - \la^*)( \phi_\la| \phi_\la) = 0. 
\ee
Hence, when $\Im \la \neq 0$, the norm of $\phi_\la$ must vanish. It means that DIM can be conceived as superpositions of an equal amount of positive and negative norm modes of the same frequency. As a result, they carry no energy. This guarantees that the growth of a DIM is compatible with energy conservation. 

In addition, $G_\la$ may have poles along the real interval $]-m; m[$. Such poles correspond to bound state modes (BSM) because they are $L^2$ on both sides. 
These modes play a crucial role in atomic physics, to determine the relativistic corrections to the energy levels of atoms~\cite{Weinberg}. 
While being mostly studied for fermionic fields, they are also relevant for the physics of bosons, 
see for instance~\cite{Greiner}. Here we see that these modes may also play a role as an intermediate step in the process of conversion of a QNM into a DIM. 

Collecting the various possibilities, using \eq{eq:phit} and evaluating the integral, one obtains the general canonical decomposition of the field in the case there is no super-radiance (since we work here with $A_0 \to 0$ for $x \to \pm \infty$)
\be \label{eq:phidecomp}
\phi(x,t)&=&\int_{\sigma_c} \lp a_\om^u e^{-i \om t} \varphi_\om^u(x)+a^v_\om e^{-i \om t} \varphi_\om^v(x) \rp \, d\om 
+ \sum_{\sigma_d \cap \mathbb R} d_a e^{-i \om_a t} \chi_a(x) \nonumber \\ 
&&+ \sum_{\sigma_d \setminus \mathbb R} \lp b_a e^{-i \la_a t} \varphi_a(x) + c_a  e^{-i \la_a^* t} \psi_a(x)\rp.
\ee
The discrete spectrum $\sigma_d$ comes from the poles of $G_\la$. As already mentioned, modes with $\Im \la \neq 0$ have a vanishing norm but the scalar product of a mode with its partner mode of complex conjugated frequency is non-zero. Instead BSM have a non-vanishing norm. (When the field is charged and quantized, they describe particles (or antiparticles) in $L^2$ bound states.) In brief, by a choice of multiplicative complex constants, the orthogonality relations for the modes belonging to the discrete sector can be taken to be
\bsub \label{eq:orthogonalityrelations}\be 
\lp \chi_a, \chi_b \rp \eg \pm \delta_{a,b}, \; \text{(BSM)}, 
\label{bsmn}\\
\lp \varphi_a, \varphi_b \rp = \lp \psi_a, \psi_b \rp \eg 0, \; \lp \varphi_a, \psi_b \rp =  \delta_{a,b} \; \text{(DIM)}.
\label{dimn}
\ee \esub

On the other hand, the continuous spectrum $\sigma_c = ]-\infty; -m] \cup [m; +\infty[$ arises from the discontinuity of $G_\la$ across the branch cut. The $u$- and the $v$- parts describe modes carrying a net momentum oriented to the right, and to the left. These can be chosen to be $in$-modes, $out$-modes, or any pair of linear superposition of these. Along the continuous spectrum, the sign of their  norm is the same as that of $\om$. Therefore the modes with $\om > 0$ correspond to particles, while the modes with $\om < 0$ correspond to anti-particles. They all carry a positive energy. To sum up the results, it is instructive to express the energy of \eq{Eelec} in terms of the coefficients entering \eq{eq:phidecomp}. Using the standard normalization 
$(\varphi_\om|\varphi_{\om'}) = \rm{sign}(\om) \delta(\om - \om')$ and Eqs.~\eqref{eq:orthogonalityrelations}, one gets 
\be
E &=& \int_m^\infty d\om \, \om \lp \abs{a_\om^u}^2 + \abs{a_\om^v}^2 + \abs{a_{-\om}^u}^2 + \abs{a_{-\om}^v}^2 \rp + \sum_a^{\rm BSM}  \om_a \abs{d_a}^2 \nonumber \\ 
&&+ \sum_a^{\rm DIM} \lp -i \la_a b_a c_a^* + h.c. \rp. 
\ee
This expression reveals that the energy is non-positive whenever unstable modes are present. 

\subsection{Outgoing boundary conditions and QNM}
\label{sub:BC}

In the preceding subsection, we saw that the localized (in the sense of footnote~\ref{lcz}) time dependent solutions of \eqref{KGE} can be expressed as a superposition of eigenmodes which are all \emph{asymptotically bounded} (plane waves or $L^2$ modes). This excludes QNM. To introduce them it is appropriate to study the {\it retarded} Green function $G^{\rm ret}_\la(x,x')$. It is defined as the analytical continuation of $G_\la$ of \eq{eq:phit} across the branch cut in the lower half-plane. This new Green function has two types of poles. In the upper complex half-plane its poles coincide with those of $G_\la$. They are thus in the spectrum as the corresponding modes satisfy \eq{eq:BC}. In the lower half-plane, however, its poles do not correspond to poles of $G_\la$. Therefore the corresponding modes violate (at least one of) the conditions of \eq{eq:BC}. Hence, they are not in the spectrum. In fact, as we explain below they are exponentially growing on both sides. These poles thus describe QNM. 

Using the conventions established after \eq{klam}; namely $\Im \la >0$, $\Im k_\la >0$,  
\eq{eq:BC} gives 
\be \label{phi1as} 
\phi_\la^{(1)}(x) &{\sim}& e^{-i k_\la x} \quad \text{ is } L^2 \text{ for } x \to -\infty , \\
\phi_\la^{(2)}(x) & {\sim}& e^{i k_\la x} \quad \text{ is } L^2 \text{ for } x\to +\infty.
\ee 
These boundary conditions correspond to \emph{outgoing} modes. Indeed, when $\la$ is close to the real axis, the imaginary part of $k_{\la}$ is related to the group velocity through
\be
k_{\la = \om + i\epsilon} = k_\om +i \epsilon \pd_\om k_\om + O \lp \epsilon^2 \rp. \label{kNRA}
\ee
For $\Im \la = \epsilon > 0$, $\Im k_{\la}$ thus has the same sign as the group velocity $v_g=(\pd_\om k_\om)^{-1}$. So, for $x \to \infty$ the wave-vector of the decaying mode has a positive group velocity: $\Im k_\la >0 \Rightarrow v_g > 0$. Similarly, for $x \to -\infty$ the decaying mode has a wave-vector $-k_\la$, and thus a negative group velocity. On the other side of the branch cut $\ep<0$, on the second sheet, the mode stays \emph{outgoing} (as it is asymptotically a superposition of waves which are analytical continuations of the outgoing ones for $\la \in \mathbb{R} \setminus [-m,m]$) but it spatially grows since $\Im(k_\la) < 0$ for $\ep < 0$. Hence it no longer satisfies \eq{eq:BC}. 

\begin{figure}[h!]
\centering
\includegraphics[width=0.9\linewidth]{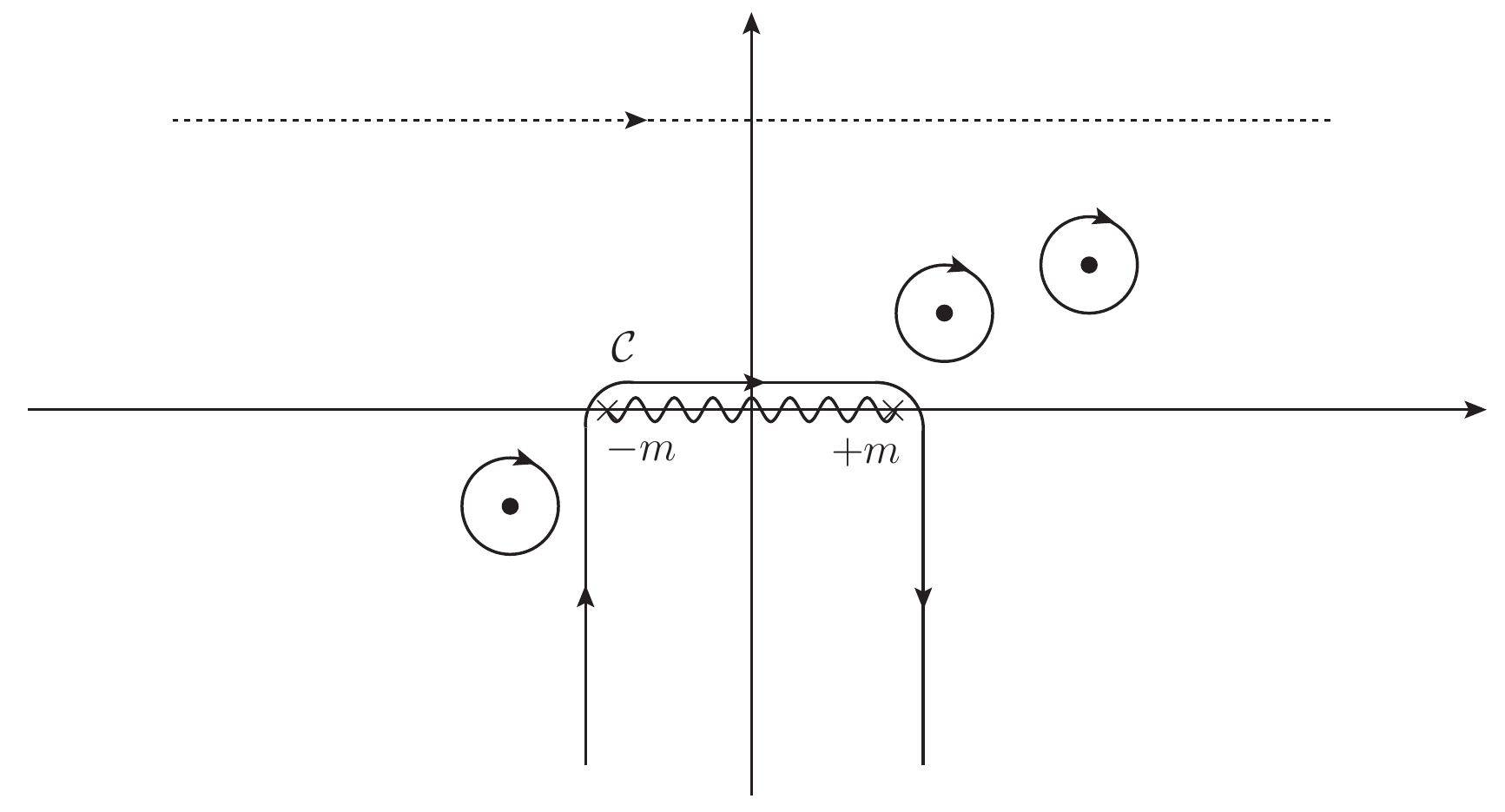}
\caption{In continuous lines we show the deformation of the retarded contour $D_{\rm ret}$ of \eq{positive_t}, shown in a dotted line, into the lower imaginary plane so as to collect the contributions of the QNM poles. The wavy line gives the branch cut and the dots are its poles. In the upper plane, we recover the poles of two DIM which also are poles of $G_\la$ of \eq{eq:phit}, see Fig.~\ref{fig:phys_contour}. In the lower plane, i.e., on the second sheet of $G_\la$, the pole of a QNM is represented.
}
\label{fig:virt_contour}
\end{figure}

In what follows, we study the analytic continuation of $G_\la$ across the branch cut. It corresponds to $G_\la^{\rm ret}$, and it agrees with the standard definition of the retarded Green function $G_\om^{\rm ret} = G_{\om + i\epsilon}$ for $\om$ real~\cite{Fulling}. To construct $G_\la^{\rm ret}$ from \eq{Gfunction}, we use \emph{outgoing} boundary condition~\footnote{To impose such boundary conditions unambiguously, the potential $A_0$ must fall off faster than $\exp(- \Im(k_\la) |x|)$. This is a rather strong condition. An alternative way to proceed is to analytically continue the scattering coefficients instead of the mode themselves. This will require $A_0$ to be an analytic function of $x$~\cite{Simon78}. In fact, for the model we shall study in section~\ref{sec:Model} the procedure will be closer to that second option. \label{ftn_potential}} instead of $L^2$. In other words, we define our Jost functions as in \eq{phi1as} for all complex values of $\la$. Doing so, the branch cut of $G_\la^{\rm ret}$ is given by that of $k_\la$, that is, the segment $]-m;m[$ (see Fig.~\ref{fig:virt_contour}).~\footnote{Notice that we could in principle have chosen a different definition of $k_\la$, giving a more complicated branch cut tracing a path in the lower half-plane from $-m$ to $m$. On top of its formal simplicity, our choice has the advantage of clearly distinguishing DIM from QNM through the sign of $\Im(\la)$. In the massless case, the branch cut of the segment $]-m;m[$ degenerates into an essential singularity at the origin (see the discussion in section 3.2 of~\cite{Berti09}). 
}
It is worth noticing that, unlike $G_\la$ of \eq{Gccsym}, the retarded Green function is not symmetric under complex conjugation, 
\be
(G_\la^{\rm ret})^* \neq G_{\la^*}^{\rm ret}.
\ee  
Hence, there is no relation between poles of the lower and upper half planes. The poles of $G_\la^{\rm ret}$ are found when the Wronskian vanishes, that is when there is a solution of the mode equation \eqref{eq:KGEs} that satisfies outgoing boundary conditions on both sides:
\be
\phi_\la(x) \underset{x \to -\infty} {\sim} e^{-i k_\la^{\rm} x} \quad \text{and} \quad \phi_\la(x) \underset{x \to+\infty} {\sim} e^{i k_\la^{\rm} x}. 
\ee
Depending on the sign of $\Im(\la)$, the outgoing modes have a different status: 
\ben
\item when $\Im(\la) > 0$, $\phi_\la$ decays on both sides, and is thus an element of $L^2$. Such poles of $G_\la^{\rm ret}$ are also poles of $G_\la$ because these coincide on the upper plane. \item when $\Im(\la) < 0$, $\phi_\la$ grows exponentially on both sides, and is thus not an eigen-vector of the spectrum. 
\een
The usefulness of QNM comes from the fact that they give the dominant contributions of the late time behavior of solutions of \eq{KGE} in the absence of DIM. To show this, we consider \eq{eq:phit} with two modifications: $G_\la$ in the integrand is replaced by $G_\la^{\rm ret}$, and the contour integration is now $D_{\rm ret}$ of \eq{positive_t} (since $D_{\rm adv}$ does not contribute for $t > 0$). Assuming that the contour can be deformed in the lower half plane as shown in Fig.~\ref{fig:virt_contour}, we collect the contributions of the poles of the QNM. 

It is also instructive to relate the poles of $G_\la^{\rm ret}$ to the scattering coefficients of the problem. When $\phi_\la^{(1)}$ satisfies outgoing boundary conditions on the left side, it decomposes on the right side as 
\be \label{eq:ScattCoef}
\phi_\la^{(1)}(x) \underset{+\infty}{\sim} \frac{1}{T(\la)} e^{-i k_\la x} + \frac{R(\la)}{T(\la)} e^{i k_\la x}, 
\ee
where $T(\om)$ and $R(\om)$ correspond  to the standard transmission and reflection coefficients for $\la = \om$ real, see \eq{scat}. One sees from \eq{eq:ScattCoef} that purely outgoing boundary conditions are satisfied when $|T(\la)| = \infty$. Therefore, poles of $G_\la^{\rm ret}$ also correspond to poles of the scattering coefficients. The poles we computed in the explicit example of section~\ref{sub:squarepotential} are exactly those of $T(\la)$, and therefore, of $G_\la^{\rm ret}$. When varying the length of the well, we saw that frequencies of QNM migrate in the complex plane and can be converted into DIM. This is allowed by the analytical structure of the Green function $G_\la^{\rm ret}$. Therefore, QNMs are not only relevant to investigate the time-dependent behavior of solutions, they also allow to understand the \emph{transition} from stable to unstable configurations. This is the main message of the present Section. In the next Section, by studying the evolution of the poles of $G_\la^{\rm ret}$ when varying an external parameter, we shall see under which precise conditions such transitions take place. 

\section{DIM and QNM from poles of the resolvent} 
\label{sec:Model}

In this section we show that the well-known model of~\cite{Friedrichs48} describing a resonance (QNM) can be generalized to describe the evolution of a QNM into a DIM under the variation of some external parameter. In this, we reveal the key role played by the relative negative sign of the norm of the trapped mode with respect to that of the continuous spectrum to which it couples. 

\subsection{Friedrichs model of a QNM}
\label{sub:FriedQNM}
We first review the model of~\cite{Friedrichs48} in its standard form, which describes the generation of a QNM as a pole of the resolvent in a second Riemann sheet. The model contains a BSM (i.e., a localized stationary solution of the linearized field equation) $|\varphi_{\om_0} \rangle$ with a positive energy which is coupled to a continuum $\left\lbrace |\phi_{\om} \rangle \right\rbrace_{\om>0}$ of real and positive frequencies. This model has the advantage of being completely solvable~\cite{Friedrichs48,Horwitz71}. It is very similar to the model of a harmonic oscillator coupled to a field, often studied to analyze decoherence or thermalization~\cite{Unruh89}.  From a field theory point of view, $|\varphi_{\om_0} \rangle$ and $\left\lbrace |\phi_{\om} \rangle \right\rbrace$ could be conceived as one-particle {\it states} living in a Hilbert space (endowed, by definition, with a (positive definite) scalar product).~\footnote{\label{nsign} Notice that in principle $|\varphi_{\om_0} \rangle$ and $\left\lbrace |\phi_{\om} \rangle \right\rbrace$  should be conceived as {\it modes} with positive norms. Yet, because of the simple dynamics of this model, the vacuum state and higher occupation number states do not mix with  one-particle states. As a result, in the present case where the norms of $\left\lvert \varphi_{\om_0} \right\rangle$ and $\left\lbrace \left\lvert \phi_{\om} \right\rangle \right\rbrace$ are all positive, it is also correct to treat these as one-particle states. To make link with the original model we may, and shall, use Dirac's notations.}

The Hilbert space of this one-particle sector may be written as the direct sum: 
\be 
\mathcal{H} = \mathcal{H}_c \oplus \mathcal{H}_d.
\ee 
$\mathcal{H}_c$ is infinite dimensional and describes the field excitations, whereas $\mathcal{H}_d \simeq \mathbb{C}$ is one-dimensional. $\mathcal{H}_c \simeq L^2(\mathbb{R}^+)$ is endowed with some scalar product $\langle \cdot | \cdot \rangle$. The field modes are normalized using a delta-function, i.e.,
\be 
\langle \phi_{\om'} | \phi_{\om} \rangle = \delta(\om-\om'), 
\ee 
where $\om$ is the real frequency of the field excitation. The atom's excitation state function $|\varphi_{\om_0} \rangle$ has a finite and normalized norm 
\be 
\langle \varphi_{\om_0} | \varphi_{\om_0} \rangle = 1. 
\label{bsm}
\ee 
The state of the total system evolves according to the linear equation 
\be \label{model_Schro}
i\pd_t |\psi \rangle = H_S \,  |\psi \rangle. 
\ee
In the Friedrich model, $H_S$ is defined by 
\be \label{eq:HFriedrichs}
H_S = \int \om' |\phi_{\om'} \rangle \langle \phi_{\om'} | d\om' + \om_0 |\varphi_{\om_0} \rangle \langle \varphi_{\om_0} | + \int V(\om') |\phi_{\om'} \rangle \langle \varphi_{\om_0} | d\om' + \int V^*(\om') |\varphi_{\om_0} \rangle \langle \phi_{\om'} | d\om' .
\ee
One easily verifies that this hamiltonian is hermitian.

The aim is now to solve the time evolution equation with the help of the stationary eigen-modes of $H_S$. In the absence of coupling, $V(\om)=0$, the eigen-modes are the above ones. When the coupling $V$ is turned on, the 
``bound'' state of \eq{bsm} disappears from the spectrum and becomes a pole of the analytical continuation of the resolvent $\hat R(\la) = (\la-H_S)^{-1}$ in a second Riemann sheet, as we now show. The operator $\hat R(\la)$ allows us to express the time-dependent solutions of \eq{model_Schro} as a contour integral 
\be \label{eq:Res_contour}
|\psi(t) \rangle = \int_{\mathcal C} e^{-i \la t} \hat R(\la) \cdot |\psi_0\rangle d\la, 
\ee
where $\mathcal C$ encircles the spectrum of $H_S$. Notice that this is the exact analog of \eq{eq:phit}. $\hat R(\la)$ is a bounded operator acting on the total Hilbert space $\mathcal{H}$ and is analytical in the parameter $\la$ except for finite poles and branch cuts. In fact, it possesses the same analytical structure as the Green function of \eq{Gfunction}. 

In this simple model, it can be shown~\cite{Horwitz71} that all the spectral properties (and the whole $S$-matrix) can be extracted from the ``reduced resolvent'' $r(\la) = \langle \varphi_{\om_0} | \hat R(\la) | \varphi_{\om_0} \rangle$ which presents the interest of being a simple scalar function of the parameter $\la$. To evaluate it, we write $\hat R(\la)$ in the unperturbed mode basis under the form
\be 
\hat R(\la)= \iint W(\om,\om') |\phi_\om \rangle \langle \phi_{\om'} | d\om d\om' + \int g_1(\om) |\phi_\om \rangle \langle \varphi_{\om_0} | d\om + \int g_2^*(\om') |\varphi_{\om_0} \rangle \langle \phi_{\om'} | d\om' + r(\la) |\varphi_{\om_0} \rangle \langle \varphi_{\om_0} | .
\ee 
We now consider the equation $\hat R(\la) \cdot (\la - H_S) = \textbf{1}$ multiplied by $\langle \varphi_{\om_0} |$ on the left and projected on $|\phi_\om \rangle$ and $|\varphi_{\om_0} \rangle$. This gives the two equations
\bsub \be 
g_2^*(\om) (\la-\om) - V^*(\om) r(\la) &=&  0 , \\
- \int g_2^*(\om) V(\om) d\om + (\la-\om_0) r(\la) &=& 1. 
\ee \esub
Combining them gives
\be \label{eq:r}
(r(\la))^{-1}= {\la - \om_0 - \int_0^\infty \frac{|V(\om)|^2}{\la - \om} d\om}.
\ee
$r(\la)$ has a branch cut for $\la \in \mathbb R^+$, giving the continuous spectrum of $H_S$. It may also have poles when $\la$ is an eigenvalue of $H_S$. Such an eigenvalue $\om_\mathbb{C}$ must satisfy 
\be 
\label{eq:omC} 
\om_\mathbb{C} = \om_0 + \int_0^\infty \frac{|V(\om)|^2}{\om_\mathbb{C}-\om} d\om.
\ee
However, by taking the imaginary part of this equation, we see that 
\be 
\Im \om_\mathbb{C} = -\Im \om_\mathbb{C} \, \int \frac{|V(\om)|^2}{(\Re \om_\mathbb{C} - \om)^2 + (\Im \om_\mathbb{C})^2} \, d \om, 
\ee
meaning that a solution of \eq{eq:omC} is necessarily real. It is nontrivial on the other hand to show that \eq{eq:omC} as no real solutions~\cite{Horwitz71}, but one easily sees that $\om_0$ itself is not a solution (unless $V$ is fine-tuned). 
Therefore, when $|V|^2$ is small enough, the bound state eigenvalue $\om_0$ is no longer in the spectrum. This is the first lesson. 

Interestingly, there is still an imprint of the bound state when one looks for poles of $r(\la)$ in the \emph{second Riemann sheet}. One can indeed analytically continue $r(\la)$ through the branch cut. For this, the function $\om \mapsto |V(\om)|^2$ must be analytically continued in the complex $\la$-plane~\footnote{This condition is analog to the condition on the potential $A_0$ as discussed in footnote~\ref{ftn_potential}.}, i.e., there must exist an analytic function $F(\la)$ such that $F(\om) = |V(\om)|^2$ for all $\om > 0$. Under this condition, one can use the standard definition $r^{\rm ret}(\la) \doteq r(\la + i\ep)$. Then using the identity $1/(x+i \ep) = \mathcal{P}(1/x) - i\pi \delta(x)$, one gets
\be \label{eq:r2ndRS}
(r^{\rm ret}(\la))^{-1} = {\la-\om_0 -\mathcal{P}  \int_0^\infty \frac{|V(\om)|^2}{\la-\om}\, d \om + 2 i \pi \Theta(-\Im(\la))F(\la)}.
\ee
By construction, this function coincides with $r(\la)$ for $\Im(\la) > 0$ and is the analog of $G_\la^{\rm ret}$. A careful check shows that, unlike $r(\la)$, $r^{\rm ret}(\la)$ is continuous across the positive real line, which is enough to ensure analyticity. The branch cut of $r^{\rm ret}(\la)$ is found along the negative real line. Indeed, the last term in \eq{eq:r2ndRS}, which is equivalent to the $i \ep$ prescription, is discontinuous on the whole real line. For $\la > 0$, this discontinuity exactly cancels that of the integral. For $\la < 0$, where the integral is continuous, it adds a new discontinuity. The same effect produces the branch cut on $]-m,m[$ in a model with a mass gap, see below subsection~\ref{subsec:massgap}, hence the different behaviours between $\la = \om \pm i \ep$ for $\om \in ]-m,m[$ mentioned in footnote~\ref{BSMvsQNM}. 

\begin{figure}[h!]
\centering
\includegraphics[width=0.45\linewidth]{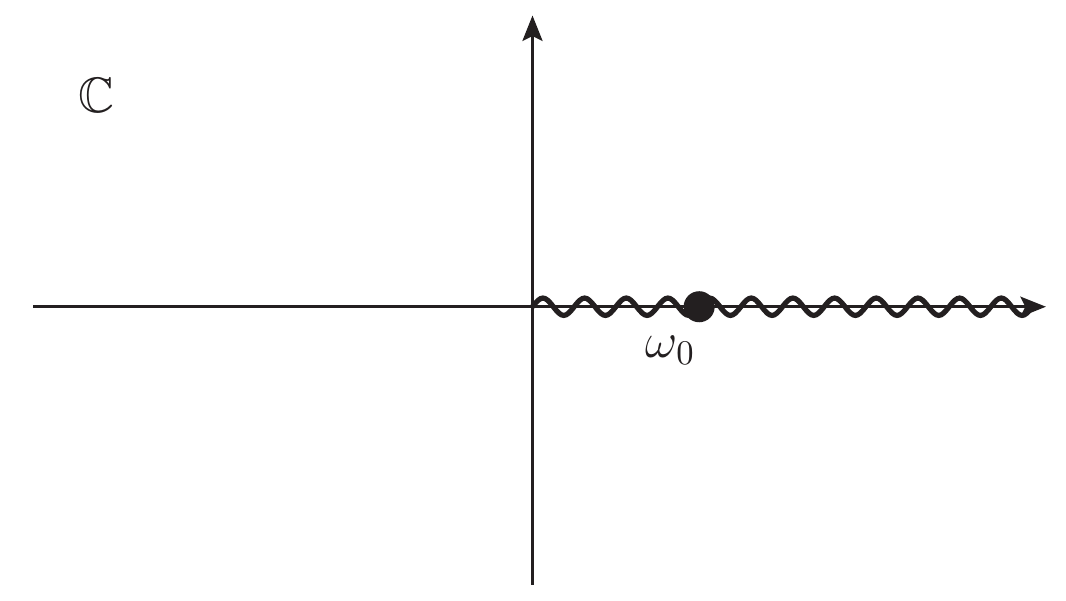}
\includegraphics[width=0.45\linewidth]{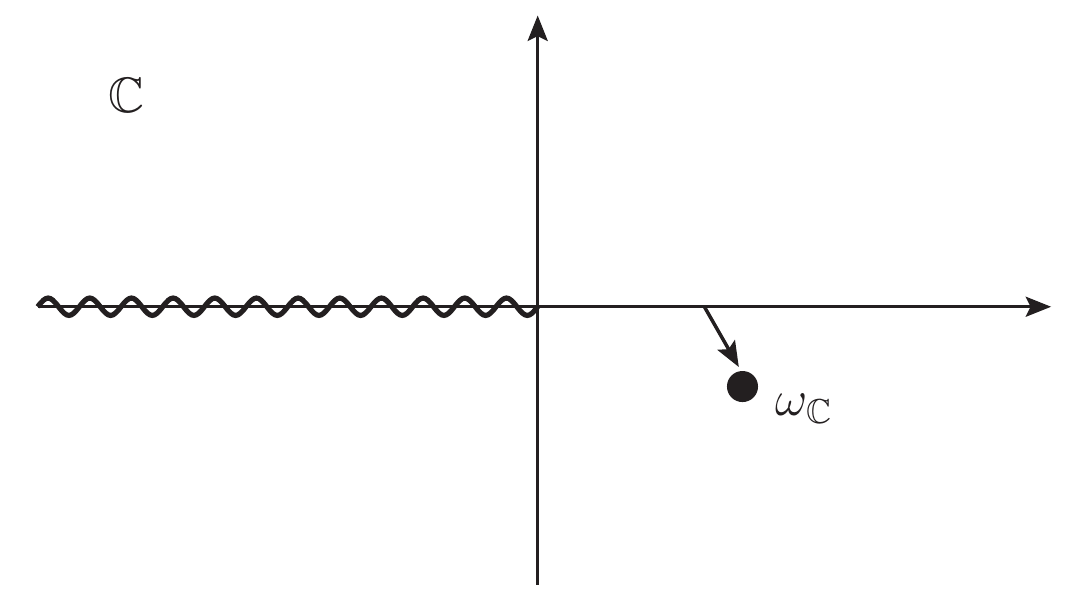}
\caption{Migration of the pole in the complex plane. On the left side, we show the branch cut of the reduced resolvent $r(\la)$ and the frequency of the bound state in the complex $\la$ plane when $V=0$. On the right side, we show the branch cut of the analytical continuation $r^{\rm ret} (\la)$ of $r(\la)$ and the displacement of the bound state frequency in the lower half-plane when turning on the interaction.}
\label{fig:QNM_spectrum}
\end{figure}

On the second sheet of $r(\la)$, for $\Im(\la) < 0$, $r^{\rm ret}(\la)$ might have poles. When $V(\om) \to 0$, there is a unique pole at $\la = \om_0 - i\ep$. When turning on the coupling, the pole $\om_\mathbb{C}$ further propagates down in the second sheet (see Fig.~\ref{fig:QNM_spectrum}). Noting $\om_\mathbb{C} = \om_0 + \delta \om + i \Gamma$, to first order in $|V|^2$, the frequency displacement and the width are given by 
\bsub \label{eq:resBWeq} 
\be
\delta \om_{\rm QNM} &=& \mathcal{P} \int \frac{|V(\om)|^2}{\om_0-\om} d\om, \\ 
\Gamma_{\rm QNM} &=& -\pi |V(\om_0)|^2.
\ee \esub 
These equations are the well-known Breit-Wigner formulas for a sharp resonance~\cite{Weinberg}. 

\subsection{Unstable Friedrichs model}
\label{sub:FriedDIM}
We now investigate what happens when the bound state possesses an \emph{opposite} norm with respect to the continuum. Therefore, we consider the same model, and only change the norm in \eq{bsm}: 
\be 
\langle \varphi_{\om_0} | \varphi_{\om_0} \rangle = -1. 
\label{negn}
\ee 
In a non-relativistic quantum mechanical model, this equation would make no sense. However such a situation arises when dealing with a bosonic field in a constant external field which is strong enough that negative energy excitations exist. As we saw in the former Section, this means that there exist some negative norm mode with positive frequency (here $\om_0 > 0$). Hence the proper way to interpret \eq{negn} is that it gives the norm of a field mode, as in \eq{bsmn}, and not that of a quantum state, see footnote~\ref{nsign}. 

\begin{figure}[h!]
\centering
\includegraphics[width=0.45\linewidth]{Figs/Free_spectrum.pdf}
\includegraphics[width=0.45\linewidth]{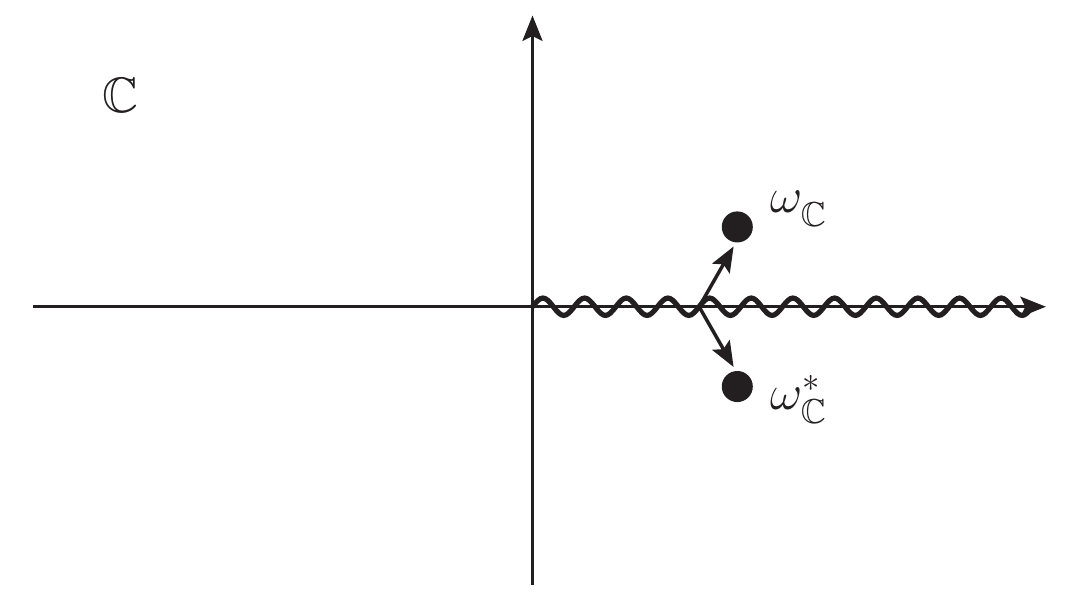}
\caption{Migration of the pole of the reduced resolvent $r(\la)$ in the complex plane. On the left side, interaction is shut down $V=0$. On the right, when $V \neq 0$, the pole splits into a pair of poles with complex conjugated frequencies which both live 
in the principal sheet.}
\label{fig:DIM_spectrum}
\end{figure}

To conserve the new (nonpositive definite) scalar product, the ``Hamiltonian'' (here defined as the generator of time translations of modes) has the form 
\be 
H_S = \int \om' |\phi_{\om'} \rangle \langle \phi_{\om'} | d\om' + \om_0 |\varphi_{\om_0} \rangle \langle \varphi_{\om_0} | + \int V(\om) |\phi_\om \rangle \langle \varphi_{\om_0} | d\om - \int V^*(\om') |\varphi_{\om_0} \rangle \langle \phi_\om | d\om' .
\ee
The only difference with respect to \eq{eq:HFriedrichs} is the minus sign in front of the third term, which is essential as it guarantees that the scalar product is conserved. Following the same construction for the reduced resolvent, we obtain  
\be 
(r(\la))^{-1} = {\la - \om_0 + \int_0^\infty \frac{|V(\om)|^2}{\la - \om} d\om}.
\ee
As before, it possesses a branch cut on $\mathbb R^+$ corresponding to the continuous spectrum. However, it may now have poles in the complex plane, and this without having to analytically continue it. Such a pole is solution of 
\be \label{eq:toyDIM}
\om_\mathbb{C} = \om_0 - \int_0^\infty \frac{|V(\om)|^2}{\om_\mathbb{C}-\om} d\om. 
\ee
Not only this equation may have solutions, but they also necessarily appear in pairs of complex conjugates, see Fig.~\ref{fig:DIM_spectrum}. 
Indeed, \eq{eq:toyDIM} is invariant under $\om_\mathbb{C} \rightarrow \om_\mathbb{C}^*$. This is reminiscent of \eq{Gccsym} and is guaranteed by the unitarity of the evolution. These two modes must be considered as a pair and not two independent modes. Indeed, the hermitian character of $H_S$ imposes the orthogonality relations 
\be
\langle \varphi_{\om_\mathbb{C}} | \varphi_{\om_\mathbb{C}} \rangle = \langle \varphi_{\om_\mathbb{C}^*} | \varphi_{\om_\mathbb{C}^*} \rangle = 0, \qquad \textrm{and} \qquad \langle \varphi_{\om_\mathbb{C}} | \varphi_{\om_\mathbb{C}^*} \rangle 
\in \mathbb{C}\setminus \{0\}.
\ee
Up to an overall factor inserted in $ | \varphi_{\om_\mathbb{C}} \rangle$ which can bring the finite overlap $\langle \varphi_{\om_\mathbb{C}} | \varphi_{\om_\mathbb{C}^*} \rangle$ to 1, these relations coincide with those we found when studying DIM in the former Section, see \eq{dimn}.  

Moreover, in our simple model, we have an explicit expression for $| \varphi_{\om_\mathbb{C}} \rangle$ in terms of the elements of the unperturbed mode basis 
\be
| \varphi_{\om_\mathbb{C}} \rangle = \int \frac{V(\om)}{\om_\mathbb{C} - \om} |\phi_\om \rangle d\om + |\varphi_{\om_0} \rangle.
\ee
Note that this construction could not have been realized in the former subsection when studying a QNM, as this mode does not belong to the spectrum. Hence, as was found when studying solutions of the Klein-Gordon equation, in the present settings a DIM is $L^2$ while a QNM corresponds to a pole in the lower half plane of the retarded Green function. In spite of this difference, when looking at the eigen-frequency perturbatively in $V$, we recover an expression very close to \eq{eq:resBWeq}. Considering for instance $\Im(\om_\mathbb{C}) > 0$, one gets 
\bsub \label{eq:unstabBWeq} 
\be
\delta \om_{\rm DIM} &=& -  \mathcal{P} 
\int \frac{|V(\om)|^2}{\om_0-\om} d\om, \\
\Gamma_{\rm DIM} &=& \pi |V(\om_0)|^2.
\ee \esub 
We see that the imaginary part has the opposite sign, and gives rise to an exponentially growing behavior in time. 

Interestingly, the above analysis can be carried out when $|\varphi_{\om_0}\rangle$ is coupled to {\it two} continua of modes with opposite norms. We define $V_+$ the coupling with the positive-norm continuum and $V_-$ the coupling with the negative-norm one, the ``bound'' state having a negative norm. We then analytically continue $r(\la)$ as in \eq{eq:r2ndRS}. Hence, a pole is an unstable mode if $\Im(\om_\mathbb{C}) > 0$ and a resonance if $\Im(\om_\mathbb{C}) < 0$. Perturbatively, we have  
\be 
\delta \om = -\mathcal{P} \int \frac{|V_+(\om)|^2-|V_-(\om)|^2}{\om_0-\om} d\om,
\ee
\be \label{eq:VpVm}
\Gamma = \pi \lp |V_+(\om_0)|^2 -|V_-(\om_0)|^2 \rp.
\ee
This shows how by changing the coupling between the mode and the two continua we can smoothly pass from a resonance when $|V_-(\om_0)| > |V_+(\om_0)|$ to an unstable mode when $|V_-(\om_0)| < |V_+(\om_0)|$. Such a transition is exactly what we observed on Fig.~\ref{fig:Roots_Elec_Massless}. 

\subsection{Modeling the mass gap}
\label{subsec:massgap}

\begin{figure}[h!]
\centering
\includegraphics[width=\linewidth]{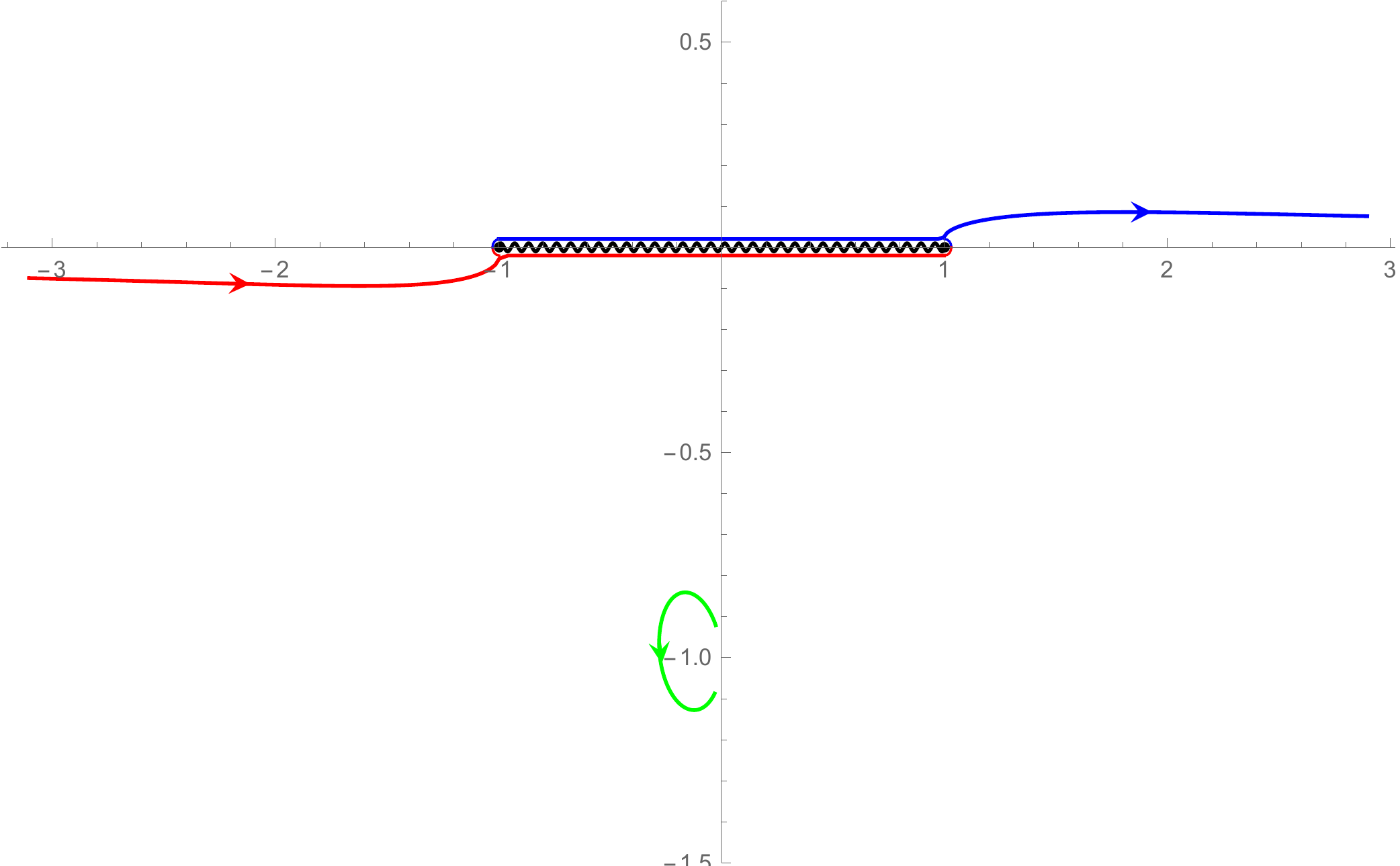} 
\caption{Trajectory of the zeros of $(r^{\rm ret}(\la))^{-1}$ given in \eq{eq:omegaf2} in the complex plane when varying $\om_0$, where $\la$ is in the units of $m$. The red (resp. blue) line lies in the lower (resp. upper) half plane. The potential $V$ is given in \eq{f} with $f_0 = 1/4$ and $m=\Lambda=1$. We notice that there is an extra QNM pole (in green). This nonperturbative pole plays no role at small couplings. At strong couplings, it may mix with the other roots and significantly affect the trajectories.}
\label{fig:Friedrichs_splitting}
\end{figure}

To obtain the behavior of Fig.~\ref{fig:Roots_Elec_Massive}, we must take into account the detailed analytic structure of the Green function. To model the mass gap, we assume that the negative continuum is defined for $\om < -m$ and the positive one for $\om>m$. Since particles and anti-particles play symmetric roles, we assume that $V_-(\om) = V_+(-\om) = V(-\om)$. Combining in a single integral the contributions of the negative and positive continua, the reduced resolvent reads 
\be \label{eq:omegaf}
(r(\la))^{-1} = \la - \om_0 + \int_m^\infty \frac{2 \om |V(\om)|^2}{\la^2 - \om^2} d\om,
\ee
To obtain both DIM and QNM, we analytically continue the integral in \eq{eq:omegaf} through the branch cut 
as in subsection~\ref{sub:FriedQNM} i.e. by the prescription $r^{\rm ret}(\la) \doteq r(\la + i\ep)$. 
We then obtain the ``retarded'' resolvent as 
\be \label{eq:omegaf2}
(r^{\rm ret}(\la))^{-1} = \la - \om_0 + 
\mathcal{P} \int_m^\infty \frac{2 \om |V(\om)|^2}{\la^2 - \om^2} d\om - 2 i \pi F(\la) \Theta \lp -\Im \la \rp.
\ee
In order to mimic the analytic structure of $G_\la^{\rm ret}$ obtained in section~\ref{sub:BC}, we choose $|V(\om)|^2 \propto \sqrt{\om^2 - m^2}$. Doing so, the function $F(\la)$ has the same branch cut as the retarded resolvent $r^{\rm ret}$, that is the segment $[-m, m]$ (just like $G_\la^{\rm ret}$ of section~\ref{sub:BC}). As we shall see, the analytic structure of $\sqrt{\om^2 - m^2}$ is the key element which explains the trajectories observed on Fig.~\ref{fig:Roots_Elec_Massive} and Fig.~\ref{fig:Friedrichs_splitting}.

The migration of the root $\la(\om_0)$ is shown on Fig.~\ref{fig:Friedrichs_splitting} for $V(\om)$ given for $\om > m$ by  
\be 
V(\om) = f_0 \frac{(\om^2 - m^2)^{1/4}}{\sqrt{\Lambda^2 + \om^2}}.
\label{f}
\ee
In this case, the trajectory can be explained perturbatively, for $|V(\om_0)|\ll m$. When $\om_0^2 > m^2$, $\la(\om_0)$  is given by \eq{eq:VpVm}, and hence it describes a DIM when $\om_0 > m$ and a QNM when $\om_0 < -m$. When $\om_0^2 < m^2$ on the contrary, $\la(\om_0)$ does not leave the real axis as $V_+(\om_0) = V_-(\om_0) = 0$. However, because of the branch cut on $[-m, m]$, there are {\it two} roots near $\om_0$, one above the cut (a BSM) and one below (a QNM). Using \eq{eq:VpVm} above and below the cut, the roots are easily obtained 
\bsub \label{eq:2split} \be
\om_{\rm BSM} &=& \om_0 -\int_m^\infty \frac{2\om |V(\om)|^2}{\om_0^2 - \om^2} d\om , \label{ombsm} \\
\om_{\rm QNM} &=& \om_0 -\int_m^\infty \frac{2\om |V(\om)|^2}{\om_0^2 - \om^2} d\om + 2 i \pi F(\om_0 - i \epsilon)
\label{omqnm} .
\ee \esub 
Notice that $2 i \pi F(\om_0 - i \epsilon)$ is real since $F(\la) \propto \sqrt{\la^2 - m^2} \in i \mathbb{R}$ for $\la \in [-m,m]$. From Eqs.~\eqref{eq:VpVm} and \eqref{eq:2split}, we perturbatively describe the trajectories of the poles in the complex plane (shown on Fig.~\ref{fig:Friedrichs_splitting}). When $\om_0 < -m$, there is a QNM, close to $\la = \om_0$. When $\om_0$ reaches $-m$, it splits into two poles with real frequencies. 
One of them propagates above the branch cut and has a frequency $\om$ given by \eq{ombsm}. The other one is below the branch cut (i.e., on the second sheet), and is thus a QNM. Both move along the branch cut and when $\om_0 = m$, they merge to give birth to a DIM. 

When taking into account nonperturbative effects in $V/m$, one obtains a slightly deformed evolution. One sees that the QNM reaches the branch cut before the value $\om_0 = -m$, 
for some $\om > -m$. It then splits into a pair of QNM with zero imaginary parts. Such QNM with infinite lifetime have also been found in rapidly rotating black holes, where they were called ``Zero Damped Modes''~\cite{Hod:2008zz,Kokkotas:2015uma}. 
One of them goes around the branch cut around $-m$, becomes a BSM, and goes to the other side of the branch cut to merge with the first one and become a DIM. This behavior is encoded in the full function $r^{\rm ret}(\la)$. It can be observed by zooming around $\pm m$ on Fig.~\ref{fig:Friedrichs_splitting}. Importantly, it was also observed on Fig.~\ref{fig:Roots_Elec_Massive}. This means that the present model captures the features observed when following the appearance of instabilities for a massive field in a deep electric well.  

\section{Case of periodic boundary conditions} 
\label{sec:discrete}

In this section we consider the model of subsection~\ref{sub:squarepotential} on a torus of length $2 \symbpiR > 2 L$. Our goal is two-fold. First, we make link with the analysis of~\cite{Fulling} by computing the eigenfrequencies on a large torus and showing in which sense the limit of an infinite one can be taken. Then, in the next subsection, we show how the coalescence of two real-frequency modes into a pair of DIM can be understood using a simple two-mode system. In Appendix~\ref{sec:degvsnondeg}, 
we further study this system by relating it to the symplectic structure of the field theory. 
This will allow us to show that a pair of DIM appears precisely when the system develops an energetic instability.

\subsection{The complex-frequency modes with periodic boundary conditions} 
\label{sub:T}

We fix $\symbpiR > L$ and impose the boundary conditions
\be 
\phi(\symbpiR,t) = \phi(-\symbpiR,t) \; \; \text{and} \; \; \pd_x \phi(\symbpiR,t) =  \pd_x \phi(-\symbpiR,t).
\ee
Since $\symbpiR > L$ the electrostatic potential is still a square well. The equation which determines the set of eigen-frequencies is now 
\be \label{eq:om'} 
\cos \left((2 \symbpiR - 2 L) k_\la+2 L k_\la^{\rm int}\right) \frac{\left(k_\la+k_\la^{\rm int}\right)^2}{k_\la k_\la^{\rm int}} - \cos \left((2 \symbpiR - 2 L) k_\la-2 L k_\la^{\rm int}\right) \frac{\left(k_\la-k_\la^{\rm int}\right)^2}{k_\la k_\la^{\rm int}}=4. 
\ee
Unlike \eq{efe}, \eq{eq:om'} is invariant under a change of sign of $k_\la$ or $k_\la^{\rm int}$, as these two wave-vectors now play symmetric roles. Concomitantly, \eq{efe} is recovered in the limit $\symbpiR \rightarrow \infty$ only in the complex upper half-plane $\Im \la > 0$. For $\Im \la < 0$, we find instead the equivalent of \eq{efe} with incoming boundary conditions. Therefore the DIM are recovered in the limit $\symbpiR \rightarrow \infty$, but in the place of the QNM we find modes with frequencies $\la_a^*$, i.e., the partners of the DIM. This limit is thus equivalent to the model of subsection~\ref{sub:squarepotential} with asymptotically bounded boundary conditions instead of outgoing ones. The reason is that QNM, due to their exponential growth at spatial infinity, can not be recovered as limits of periodic modes. A related fact is that the Green function $G_\la$ has no branch cut as the spectrum is discrete. Instead, it has poles on the two half-lines $\la \in (-\infty, -m]$ and $\la \in [m, \infty)$, with a density growing linearly in $\symbpiR$.  In the limit $\symbpiR \to \infty$, the accumulating poles become the branch cut of \Fig{fig:phys_contour}. For finite values of $\symbpiR$ there is thus no clear analog of the retarded Green function which could be used to define QNM. 
(Notice however that their imprint on the scattering coefficients subsists. Indeed, when $\symbpiR$ is finite but large, the three families of peaks discussed at the end of subsubsection~\ref{subsub:modessquarewell} remain, although the third one does not correspond to eigen-modes on the torus.)

Let us now briefly study  \eq{eq:om'} per se, because it illustrates the disappearance of a DIM, something that did not occur in Section~\ref{sec:KG}. For $L=0$, the discrete solutions in $\la$ are all real. When increasing $L$, these eigenfrequencies move along the real axis and eventually frequencies corresponding to modes with opposite norms cross each other provided $e A > 2 m$. We observe that this crossing gives birth to pairs of DIM. When keeping increasing $L$ the imaginary parts of these frequencies increase before going back to zero. Indeed, when $L$ reaches $\symbpiR$, all eigenfrequencies are again real, in virtue of the symmetry $k_\la \to k_\la^{\rm int}$, $L \to (\symbpiR - L)$. 
Interestingly, a direct calculation shows that at least one pair of complex eigenfrequencies exists for $L=\symbpiR/2$ provided $|e A| > 2 m$, however small $\symbpiR$ is. 

\begin{figure}[h!]
\begin{center}
\includegraphics[width=0.49 \linewidth]{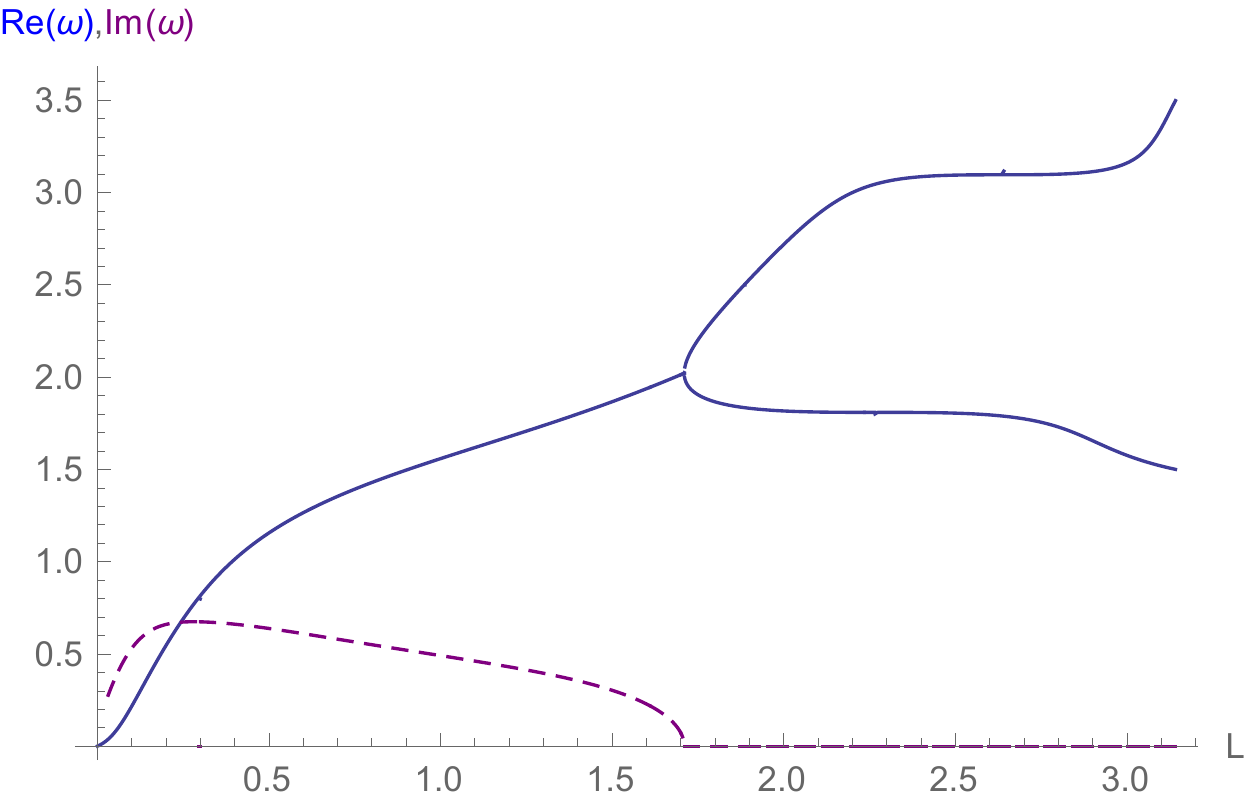}
\includegraphics[width=0.49 \linewidth]{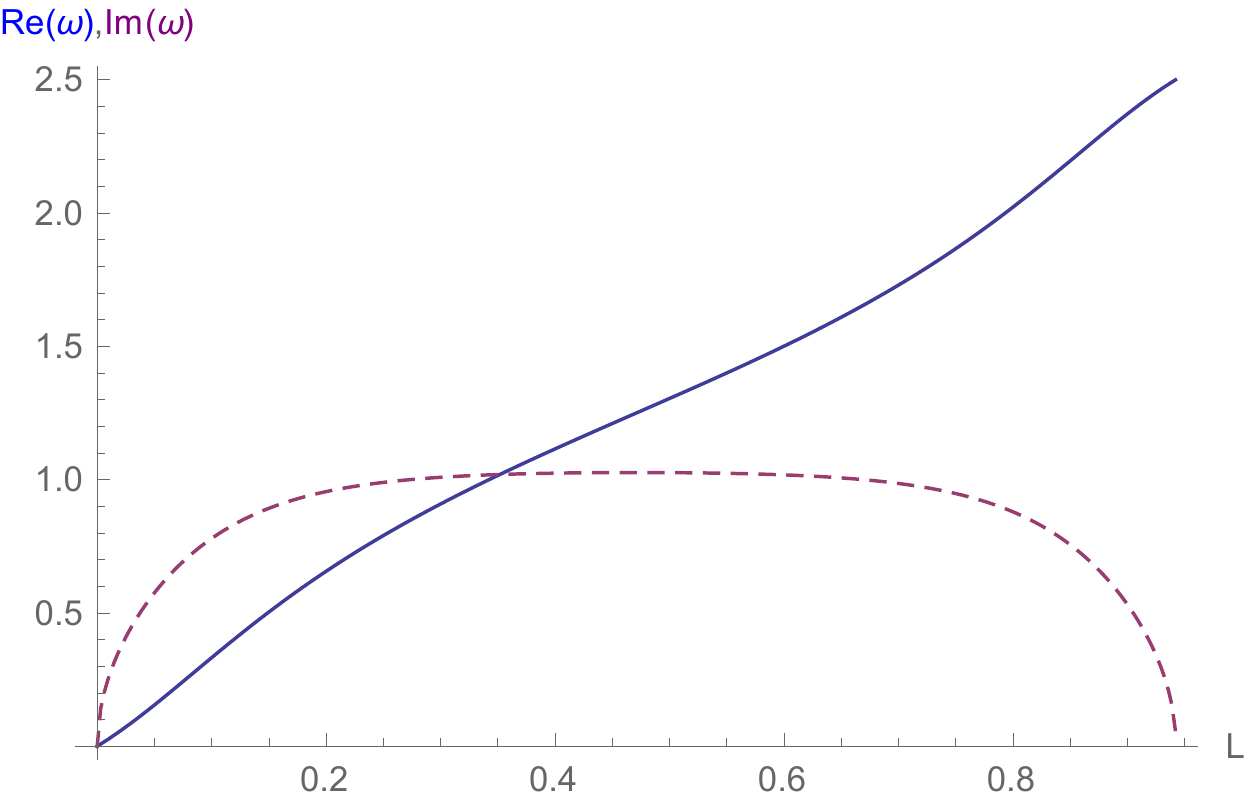}
\caption{Real (blue, solid) and imaginary (purple, dashed) parts of the first complex eigenfrequency as functions of $L$, for the massless electric case with periodic boundary conditions. These two plots are made with the same value of $e A = 2.5$ but different values of $\symbpiR$: $\symbpiR=\pi$ (left) and $\symbpiR=0.3 \pi$ (right). While their growth is similar for very small values of $L$, on the left panel, the DIM disappears around $L \sim \symbpiR/2$, while on the right one 
it disappears only when $L$ reaches $\symbpiR$, following a curve which is the symmetric of that characterizing its birth. 
} \label{Fig:om_T} 
\end{center}
\end{figure}

In the massless case, as in Section~\ref{sec:KG}, the first DIM appears as soon as $L \neq 0$, see Fig.~\ref{Fig:om_T}. When $e A \symbpiR \gg 1$, 
it disappears when $L$ reaches a critical value as can be seen on the left plot. In this case, it gives rise to two real frequency modes with opposite norms. Instead, when $e A \symbpiR \lesssim 2 \pi$, the first DIM only disappears when $L \to \symbpiR$, as shown on the right plot. This example clearly reveals the importance of the length $\symbpiR$ 
expressed in units of $(eA)^{-1}$, the inverse length fixed by the external potential. (The evolution of eigen-frequencies of a massive field is more complex because the inverse mass provides another length scale.) 

\subsection{Mode coalescence and birth of DIM}
\label{aac}

In standard quantum mechanics, it is well known that a small coupling between two states becomes very important when the unperturbed frequencies cross each other when varying some external parameter. Indeed, one finds that the frequencies of the true eigenstates do not cross each other, in a phenomenon called \emph{avoided crossing}~\cite{Gottfried}. Following the same approach as in subsection~\ref{sub:FriedDIM}, we study below the evolution of a two-mode system under some external variation when the two modes have opposite norms. We shall see that the main results of the former subsection, namely that eigen-modes of opposite norms can coalesce and form a DIM when varying an external parameter, are recovered using this simple formalism. In fact, this is the basic process generating DIM in the discrete case (see Fig.~\ref{Fig:om_T}). 
This formation was observed in a variety of contexts, two prominent examples being the Klein-Gordon equation in an electric field~\cite{Fulling,Fulling76,Schiff40} and the Gross-Pitaevski equation in Bose-Einstein condensates~\cite{Jackson,Nakamura}. Here we give the basic elements responsible for this phenomenon.

\begin{figure}[h!]
\begin{center}
\includegraphics[width=0.5 \linewidth]{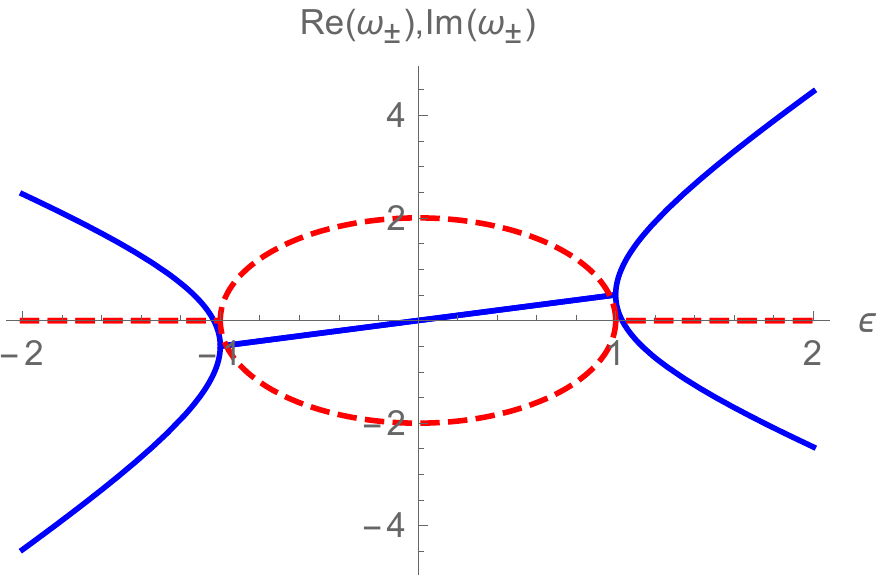} 
\caption{Real (Blue, plain) and imaginary (Red, dashed) parts of the two eigenvalues of \eq{true} as functions of $\epsilon = (\om_2 - \om_1) / 2 \in [-2,2]$ for fixed $\beta=1$ and for $\om_1 = 3 \ep /2$, $\om_2 = - \ep / 2$. As long as $|\epsilon| > \beta=1$, the imaginary parts of the two eigenfrequencies $\om_\pm$ vanish. When $|\epsilon| < 1$ instead, these imaginary parts no longer vanish and are opposite to each other. 
}\label{fig:lvlcross_comp}
\end{center}
\end{figure}

To make contact with the Klein-Gordon equation \eqref{KGE}, we work in the restricted set of solutions of the form 
\be
\phi(t,x) = a(t) \phi_{\om_1}(x) + b(t) \phi_{\om_2}(x),
\ee
where $\phi_{\om_1}$ and $\phi_{\om_2}$ are two eigen-modes at a specific value of the background parameter. Then $\Phi \doteq (a,b)^T$  
satisfies an evolution equation of the form 
\be \label{eq:S}
i \partial_t \Phi = \symbHS \cdot \Phi,
\ee
with the Hamiltonian 
\be \label{eq:2X2HS}  
\symbHS = \bmat \om_1 &  \beta \\ -\beta^* & \om_2 \emat,
\ee 
where $(\om_1, \om_2) \in \mathbb{R}^2$, and where the complex number $\beta$ governs the interaction between the unperturbed modes $\phi_{\om_1}$ and $\phi_{\om_2}$. In the present case, we study the interaction of two modes of \emph{opposite} norms. One verifies that the signs of the nondiagonal elements of $\symbHS$ ensures that the norm 
\be 
(\Phi, \Phi) = \Phi^\dagger 
\begin{pmatrix}
1 & 0 \\ 
0 & -1
\end{pmatrix} 
\Phi
\ee
is conserved in time. The eigenvalues of $\symbHS$ are
\be 
\omega_\pm  = \frac{\om_1 + \om_2}{2} \pm \sqrt{(\om_1 - \om_2)^2 - 4|\beta|^2}.
\label{true}
\ee
In the absence of coupling, $\beta = 0$, the time evolution is given by $a = e^{-i\om_1 t}$ and $b = e^{-i\om_2 t}$. If they depend on some external parameter, $\om_1$ and $\om_2$ may then cross each other without interfering. However, when the modes are not independent (which is generally the case, unless the matrix element $\beta = \langle 1 | \symbHS | 2 \rangle$ exactly vanishes), the eigen-frequencies of \eq{true} no longer cross each other but become complex conjugated of each other, as shown on Fig.~\ref{fig:lvlcross_comp}. Hence the two-mode system gives rise to a DIM and its partner mode. 

Let us see in closer details how this occurs. If $2 \left|\beta \right| < \left| \om_1 - \om_2 \right|$, the eigen-values of \eq{true} are still both real. When $2 \left|\beta \right| = \left| \om_1 - \om_2 \right|$, they become equal to each other. Further increasing $\left|\beta \right|$ gives two complex-conjugate eigenvalues. This phenomenon is very similar to the \emph{avoided crossing}~\cite{Gottfried}. The only difference is in the relative sign of the norms of the two modes involved. When they are equal, the eigen-frequencies are repelled from each other, hence the avoided crossing. When their sign is opposite, the two frequencies merge and acquire opposite imaginary parts. This is exactly what we obtained in subsection~\ref{sub:FriedDIM}.  

\begin{figure}[h!]
\begin{center}
\includegraphics[width=0.5 \linewidth]{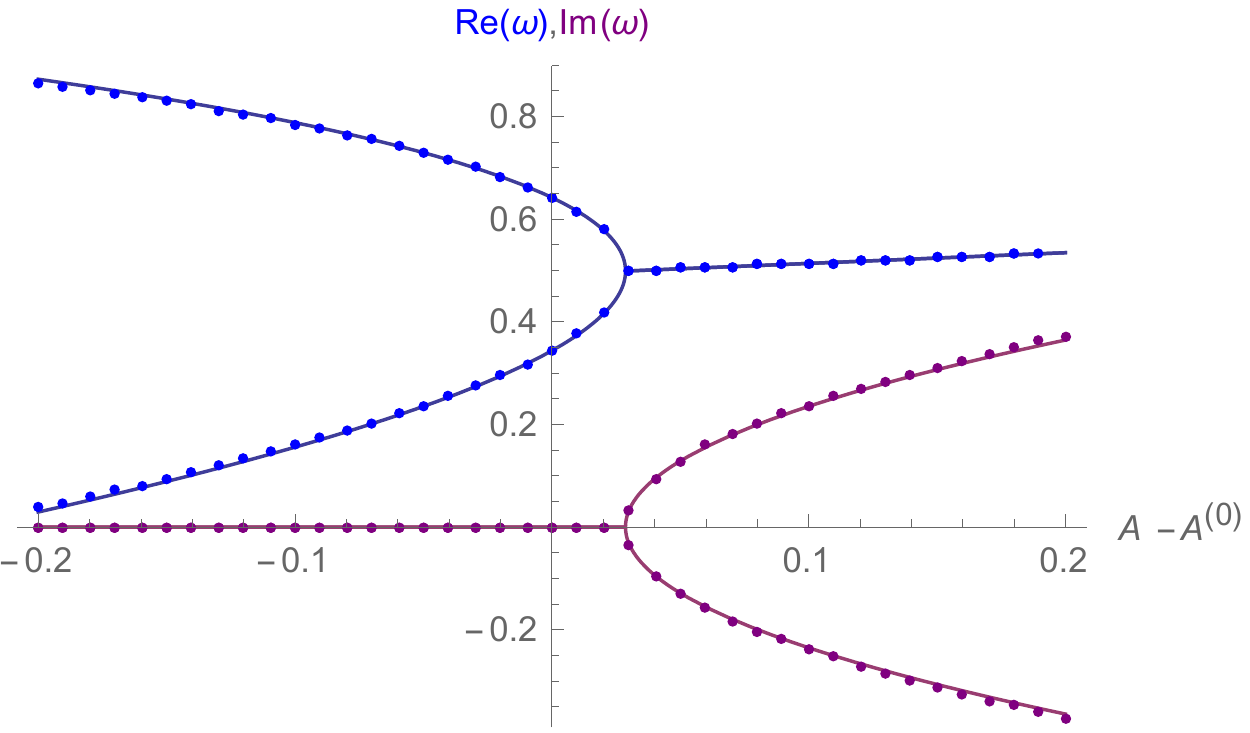} 
\caption{Real (blue) and Imaginary (purple) parts of the two first complex eigenfrequencies as functions of $A-A^{(0)}$, and for $\symbpiR = 2 \pi/10$, $L=0.06$, $m=1.0$, and $A^{(0)}=2.5$. Plain lines are computed using the 2 by 2 model where the two unperturbed modes are eigen-modes of the Hamiltonian for $A=A^{(0)}$. Dots are the exact eigenfrequencies, obtained by solving \eq{eq:om'}. One barely sees that the plain lines start differing from the dots for $A-A^{(0)} \sim \pm 0.2$.  This agreement indicates that the 2 by 2 model accurately describes the coalescence of two modes with opposite norms.
}\label{fig:comp_2by2}
\end{center}
\end{figure}

This $2 \times 2$ model is more than a formal analogy. Indeed, as we now show, it is able to accurately describe the merging of two real eigenvalues when dealing with a field theory. As an example, we consider the solutions of the Klein-Gordon equation under some variation of the electric potential $A^{(0)}$. We work with the model described by \eq{eq:om'} and we choose two eigen-modes $\phi_1$, $\phi_2$ with nearby frequencies and opposite norms for a given value $A^{(0)}$ of $A$. Explicitly we work with the following values of the parameters: $eA^{(0)} = 2.5$, $L = 0.06$, and $\symbpiR = 2 \pi / 10$. We then compute numerically the $2 \times 2$ matrix $(\phi_i | i \pd_t \phi_j)$, 
$i,j \in \left\lbrace 1,2 \right\rbrace$ for different values of $A$, giving the $2 \times 2$ restriction of $H_s$ of \eq{eq:2X2HS}. Since $\phi_1$ and $\phi_2$ are eigenmodes only for $A = A^{(0)}$, the coefficient $\beta$ is non-vanishing when $A \neq A^{(0)}$. The approximation then consists in describing the evolution of the two eigenfrequencies by discarding all the other modes. In \Fig{fig:comp_2by2}, we compare the results so obtained with the exact ones obtained numerically when taking all modes into account. The smallness of the relative deviations, which are of order $10^{-4}$ for $A - A^{(0)} \approx 0.1$, confirms that the $2 \times 2$ model quantitatively describes the mode merging giving birth to a pair of DIM. The lesson 
is the following: even though we are studying a field with an infinite {\it discrete} number of degrees of freedom, each mode merging is, in a neighborhood of the merging, effectively described by a $2 \times 2$ model. 

\section{Conclusions}
\label{Concl} 
This paper presents a study of the birth of instabilities when a bosonic scalar field is coupled to a strong external field. 
To perform the analysis, we consider a charged scalar field in 1+1 dimension in an external electric potential. We first analyze 
analytically solvable examples, i.e., 
single-step and square well potentials. In the first case, when the height of the step is large enough, there is a finite frequency interval in which stationary modes with opposite norms mix with each other. As a result, there is an energetic instability and a steady mode amplification (pair creation in quantized settings) which conserves the total energy and charge. In the second case, for the same height of the step, the continuum spectrum of scattered modes is replaced by a discrete set of modes (called DIM) which appear in pairs with complex conjugated frequencies, and which encode the dynamical instabilities of the system. On very general grounds, we showed that these modes are asymptotically square integrable, have a vanishing norm (i.e., carry no electric charge) and carry no energy. We observe a close relation between QNM and DIM as QNM get smoothly converted into DIM as one varies external parameters. For massive fields, this conversion occurs with an intermediate step, where QNM first become bound state modes (BSM) before becoming DIM. 

We then present a general resolution of the Klein-Gordon equation in an electric field in order to establish the key differences and relations between QNM and DIM (Sec.~\ref{sub:modebasis}). We show that DIM are associated with complex eigen-values of the equation, while QNM only show up by properly analytically continuing the Green function in a second Riemann sheet, i.e. accross the branch cut provided by the real (continuous) spectrum. This continuation is equivalent to a change of boundary conditions (Sec.~\ref{sub:BC}). As we explain, when working with \emph{outgoing boundary conditions}, DIM, BSM, and QNM are treated on an equal footing, and one can study the transition from one to the other. This reveals a new role of QNM, which are not only relevant to understand the time-dependent ring down of the field, but also the transition from a stable to unstable configuration when the background varies. Moreover, when the mass is nonzero, we observe that when a QNM is converted into a BSM, another outgoing mode is generated. Although its frequency has a vanishing imaginary part (it is ``infinitely long-lived''), it describes a QNM, as it does not belong to the discrete spectrum. In the presence of a mass gap, this QNM plays a crucial role in the parametric transition from a stable to an unstable system. To understand the various transitions QNM $\to$ BSM $\to$ DIM, we present a model which illustrates the minimum ingredients which are necessary for such a transition to occur (Sec.~\ref{sec:Model}). This model is a generalization of the Friedrichs model. In its original version, the model is defined in a (positive definite) Hilbert space, and describes the appearance of a resonance (i.e.,  a QNM in our language) in the lower half plane of the retarded Green function when coupling a localized isolated degree of freedom to a continuum of modes. As we show, if the isolated degree of freedom is coupled to a continuum of \emph{opposite} norm, it gives rise to a DIM instead. Hence, when coupled to two continua with a different sign of norm, the degree of freedom can smoothly transit from a QNM to a DIM by varying the relative coupling. Moreover, by imitating the analytic structure of a mass gap, we show that such a simple model reproduces the complicated behavior of the massive case: the transition QNM $\to$ BSM $\to$ DIM, with the appearance of a real-frequency QNM as partner of the BSM. 

To complement this analysis of the possible appearance of instabilities, we consider the case of periodic boundary conditions (Sec.~\ref{sec:discrete}). In this case the mode spectrum is necessarily discrete. As a result, there seems to be no proper way to define QNM. Instead, DIM can still exist. As we show, their appearance results from a merging of two real frequency modes with opposite norm. This is the pendant of the standard avoided crossing of quantum mechanics when the scalar product is not positive definite. To complete the case of a discrete spectrum, we compare in App.~\ref{sec:degvsnondeg} this case to that of a real (uncharged) scalar field in stable/unstable backgrounds. We show that there are in general two classes of DIM. In the first one, which is the most frequent, the DIM is described by a complex degree of freedom, and a complex eigen-frequency. In the second case, the DIM is described by a real degree of freedom with two eigen-frequencies, which are purely imaginary and opposite to each other. A typical example of this in black hole physics is the Gregory-Laflamme instability~\cite{Gregory93,Gregory:2011kh}, but this case is also found in hydrodynamical flows (see for instance \cite{doi:10.1146/annurev.fl.12.010180.001511,dias1996mathematical}) and in the black hole laser effect. 

In brief, our work presents a classification of the possible linear dynamical instabilities one might encounter in stationary backgrounds. It reveals the key roles of the conserved scalar product, the analytic structure of Green functions, boundary conditions, and QNM. 

\begin{acknowledgments}
A.C. would like to thank Christian G\'erard and J\'er\'emie Joudioux for several discussions concerning mathematical aspects of this work.  
\end{acknowledgments}

\appendix 

\section{Symplectic structure, energetic instabilities, and DIM} 
\label{sec:degvsnondeg}

In this appendix we show the key role played by the symplectic structure of the field theory when  instabilities appear. This will allow us to establish the relationship between the (real) energies of field configurations and the (complex) frequencies of DIM of the same field theory. In particular, we shall prove that for systems described by a discrete number of degrees of freedom, DIM and energetic instabilities (EI) occur for the same value of the parameters describing the external potential. The symplectic structure will also allow us to explain the fact that there exist two qualitatively different types of DIM, which we call degenerate and nondegenerate. Nondegenerate DIM, as the ones appearing in the model of subsection~\ref{sub:modebasis}, involve one \textit{complex} degree of freedom. Their frequencies are in general complex. A degenerate DIM instead, as the Gregory-Laflamme instability \cite{Gregory93,Gregory:2011kh}, 
involves only a single \textit{real} degree of freedom and has a purely imaginary frequency. These two types of DIM have been observed in \cite{MichelParentani} when further analyzing the black hole laser effect~\cite{Corley99,Leonhardt08,Coutant10,Finazzi10}. 

In the following we first recall the results of \cite{MichelParentani}, calling for a better understanding of DIM and their relations with EI. We then rephrase the model of section~\ref{aac} in a Hamiltonian language which delivers the relationships we are looking for.

\subsection{Black hole laser instabilities}
\label{bhli}

In \cite{MichelParentani,Michel:2015pra}, it was shown that nondegenerate 
 DIM appear through a non-trivial three-step process~\footnote{These references focused on the case where the parameters in the two asymptotic regions are equal. Then the second and third steps described below occur for the same value of $L$.}. The set of DIM was studied when increasing the length $2 L$ of the supersonic region. For small values of $L$, there is no DIM but only QNM. For increasing $L$, DIM appear as follows 
\begin{itemize}
\item a QNM turns into a 
degenerate DIM with a purely imaginary frequency for $L > L_1$;  
\item a second degenerate 
purely imaginary eigenfrequency appears in the same way for $ L > L_2 > L_1$;  
\item these two frequencies merge into a complex 
nondegenerate one for $ L > L_3 > L_2$.  
\end{itemize}
Importantly, when studying nonlinear stationary solutions of the Gross-Pitaevskii equation, it was also found that new {\it energetic instabilities} (EI) appear precisely at $L_1$ and $L_2$ but not at $L_3$. This indicates that in the present case, there is a direct correspondence between DIM and EI. In addition, this process is repeated periodically, increasing the number of DIM and EI each time $L$ crosses a new critical value. It is also crucial to note that the DIM for $L<L_3$ and $L>L_3$ are qualitatively different and that only the second one is directly related to Hawking radiation. Indeed, the DIM for $L<L_3$ only has a \textit{real} degree of freedom, and the eigen-frequency is purely imaginary.

\subsection{Symplectic structure, EI and DIM}
\label{bhli1}

We now present a model which allows to understand the above observations. As in subsection~\ref{aac}, it is based on restriction  to a finite-dimensional sub-space of modes. A restricted Hamiltonian can then be defined to determine the time-evolution of their coefficients. To be more precise, we consider solutions of a scalar, linear field equation in the form 
\be \label{eq:fewmodes}
\phi(x,t) = \sum_{i = 1}^N q_i(t) \, \phi_i(x), 
\ee 
where $\phi_i$ are known orthogonal and normalized (in the sense of the $L^2$ scalar product) functions. Our goal is to study the evolution of the coefficients $q_i$ when taking into account only the interactions between the $N$ modes appearing in \eq{eq:fewmodes}. 

In the following, we work with {\it real} time dependent coefficients. (This analysis extends to the case of complex ones after decomposing them into real and imaginary parts, see the next subsection.) We then exploit the fact that the field equation under study has a canonical Hamiltonian structure. Plugging \eq{eq:fewmodes} into the expression of the Lagrangian one can define the conjugate moments $p_i$ of the $q_i$. 

Let us first consider the case with two real degrees of freedom $q_1, q_2$. We define the vector 
\be \label{eq:X} 
X \equiv 
\left(
\begin{array}{cc}
q_1 \\
q_2 \\
p_1 \\
p_2
\end{array}
\right).
\ee
As the field equation is linear, the Hamiltonian is quadratic in $(q_1,q_2,p_1,p_2)$:
\be 
H = \frac{1}{2} X^T \symbHE X,
\label{H1}
\ee
where the hamiltonian matrix $\symbHE$ is real and symmetric. 
$H$ is the energy of the mode, not to be confused with the time-translation operator $\symbHS$. The eigenvector of $\symbHE$ corresponding to the eigenvalue $E_i$ gives initial conditions for a mode of energy $E_i$ via \eq{eq:iniphi}. 
Let us write the Hamilton equations 
\be 
\left\lbrace
\begin{array}{ll}
 \frac{dq_i}{dt}=\frac{\partial H}{\partial p_i} &   \\
 \frac{dp_i}{dt}=-\frac{\partial H}{\partial q_i} &  
\end{array}
\right. ,
\ee
in a matrix form 
\be 
i\frac{d}{dt}\left(
\begin{array}{c}
 q \\
 p
\end{array}
\right)=i \left(
\begin{array}{cc}
 0 & \mathbf{1} \\
 -\mathbf{1} & 0
\end{array}
\right)\symbHE \left(
\begin{array}{c}
 q \\
 p
\end{array}
\right)= \symbHS
\lp
\begin{array}{c}
 q \\
 p
\end{array}
\rp .
\label{Schreq}
\ee
In this equation, $\mathbf{1}$ is the $2$ by $2$ identity matrix, and 
\be 
J\equiv \left( \begin{array}{cc}
 0 & \mathbf{1} \\
 -\mathbf{1} & 0
\end{array} \right), 
\ee
is the symplectic matrix. Indeed, the Poisson brackets between $q_i$ and $p_j$ are all encoded in \be \label{eq:scalXY}
\left\lbrace X,Y \right\rbrace \equiv X^T \, J \, Y.
\ee
This bilinear form should be conceived (up to a constant prefactor) as the restriction of the scalar product of \eq{Ksp} to the present subspace. Indeed, when considering two solutions of~\eq{KGE} given by the initial conditions
\be \label{eq:iniphi}
\phi^{(a)}(x,t=0) = \sum_i q_i^{(a)} \phi_i(x), \quad \pi^{(a)}(x,t=0) = \sum_i p_i^{(a)} \phi_i(x), \\ 
\phi^{(b)}(x,t=0) = \sum_i q_i^{(b)} \phi_i(x), \quad \pi^{(b)}(x,t=0) = \sum_i p_i^{(b)} \phi_i(x),
\ee
where $\ep_i = \pm 1$ is the sign of the norm of the $i$th mode, introduced for practical convenience, we have
\be 
(\phi^{(a)},\phi^{(b)}) = \sum_{i,j} (q_i^{(a)} p_j^{(b)} - q_j^{(b)} p_i^{(a)}) \times i \int \phi_i^* \phi_j \, dx = i \sum_i (q_i^{(a)} p_i^{(b)} - q_i^{(b)} p_i^{(a)}).
\ee
The right hand side of the last equation is (up to the factor $i$) \eq{eq:scalXY}, where $X$ is defined by \eq{eq:X} for the $(a)$ variables, and $Y$ is the same vector with the $(b)$ ones. 

Because $J^2 = - 1$, the matrix $\symbHS$ of \eq{Schreq}
\be \label{eq:HEtoHS}
\symbHS \doteq i J \symbHE,
\ee 
gives back the Hamiltonian $H$ of \eq{H1} when using the symplectic scalar product of \eq{eq:scalXY} 
\be 
H = \frac{i}{2} \, \left\lbrace X,\symbHS \, X \right\rbrace .
\ee  
This identity is the restriction in the $(q_1,q_2)$ subspace of the relation obeyed by a complex scalar field:  
\be
H[\phi] = (\phi, i\pd_t \phi), 
\ee
between the field Hamiltonian $H[\phi]$ and the scalar product in a quadratic field theory, as mentioned after \eq{Eelec}. Is it interesting to note that although the norm $(\phi,\phi)$ identically vanishes for any real field configuration, the energy $\propto (\phi, i\pd_t \phi)$ does not, and correctly differentiates positive and negative energy waves, see footnote~\ref{waveen}. 

\begin{figure}[h!]
\begin{center}
\includegraphics[width=0.49 \linewidth]{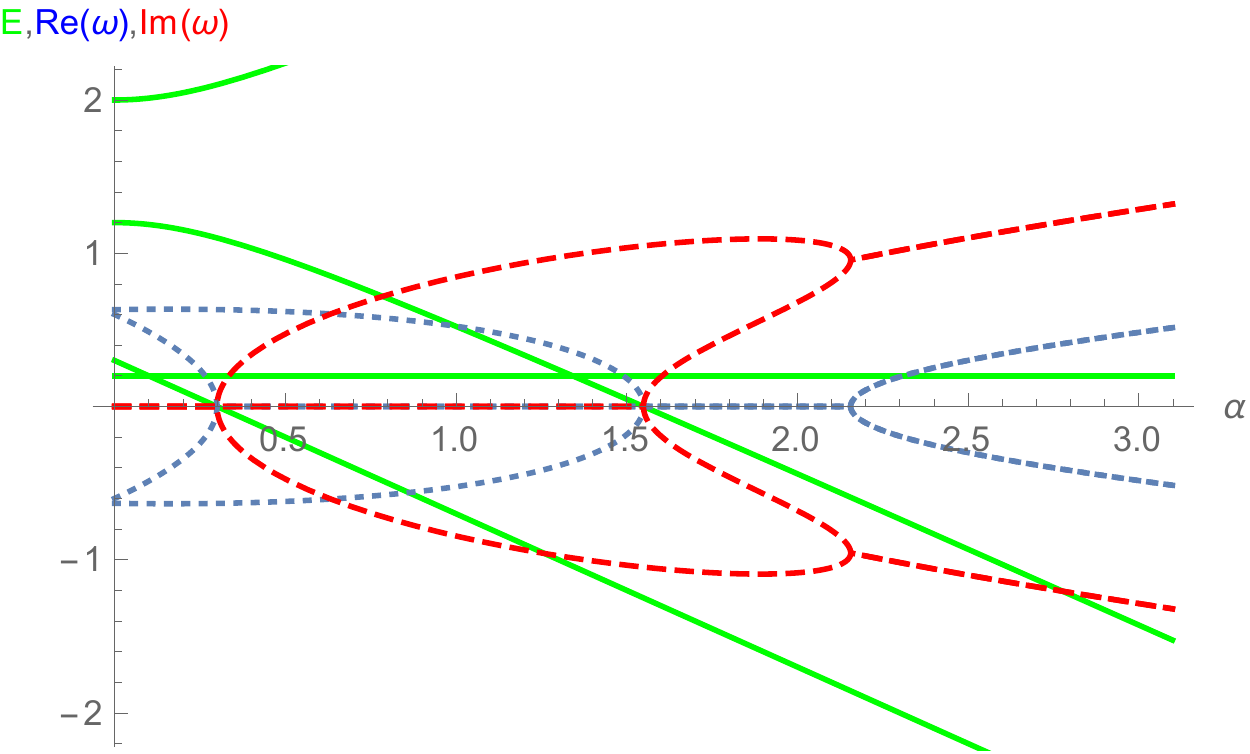} 
\includegraphics[width=0.49 \linewidth]{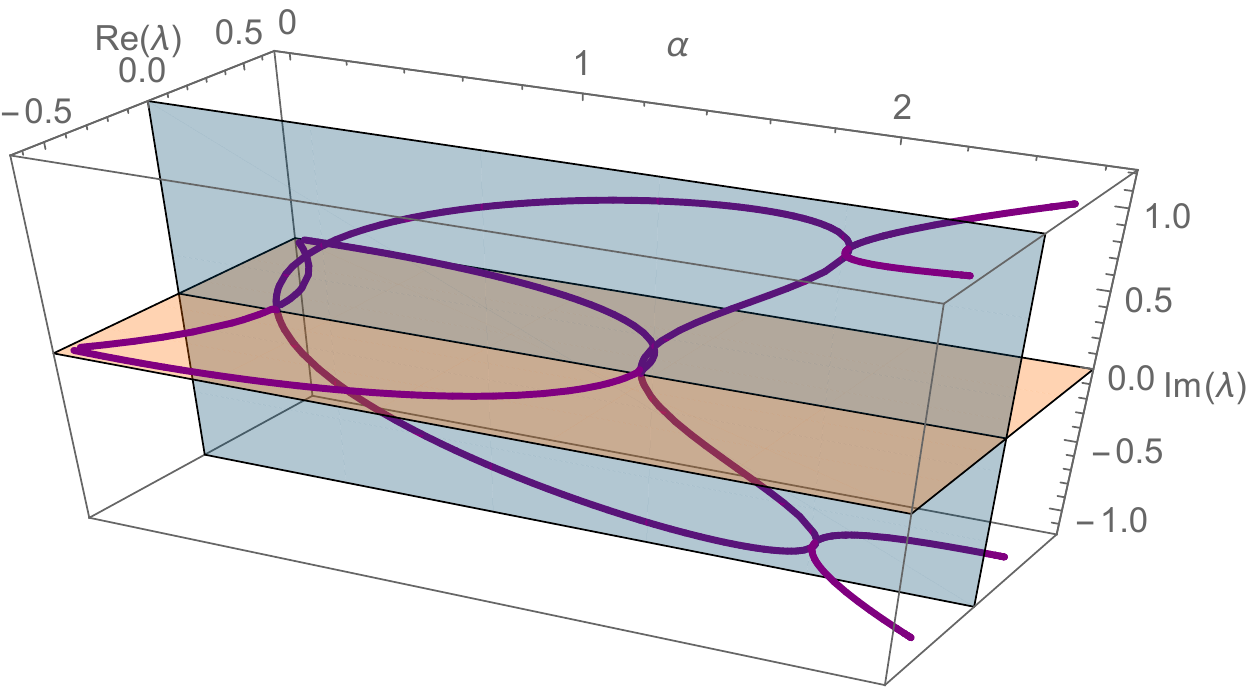} 
\end{center}
\caption{Left panel: The green, continuous lines represent the four eigenvalues of the Hamiltonian matrix $\symbHE$ when $\symbHE$ has the form (\ref{eq:Hqq}), with $E_{p_2} = 0.3-\alpha$, $E_{q_1} = 2.0, E_{q_2} = 1.2$, and $E_{p_1} = 0.2$. These real numbers give the energy of the corresponding real field configurations. The blue, dotted (red, dashed) lines gives the real (imaginary) parts of the four eigenvalues of $\symbHS$. (They are multiplied by 2 to make the plot clearer.) These complex numbers give the eigen-frequencies of the complex modes $\propto e^{- i \la t}$. We observe that two degenerate DIM appear at the same time as the two energetic instabilities at $\alpha \approx 0.3$ and $\alpha \approx 1.55$. We also observe that they merge to give a nondegenerate DIM for $\alpha \approx 2.16$. Right panel: We show the 4 eigenvalues in the complex plane as functions of $\alpha$. The orange plane corresponds to $\Im \la = 0$ and the blue one to $\Re \la = 0$. The 4 purple lines show the positions of the 4 eigenvalues in the complex plane. 
}\label{Fig:omBHL}
\end{figure}

Since $\symbHE$ is a symmetric real matrix, its eigenvalues are necessarily real. 
A negative eigenvalue of $\symbHE$ indicates that there is an EI. In that case the energy of the system is not bounded from below (in the absence of higher-order terms in the field amplitude in the Hamiltonian). Interestingly, the eigenvalues of $H_S$ may be complex. In particular, there is a DIM when a pair of eigenvalues of $\symbHS$ are complex. Since in full generality the spectrum is invariant under complex-conjugation,  as discussed in the main text, complex eigenvalues only arise in pairs. (In the present settings, it can be traced to the fact that $\symbHS$ is hermitian for the symplectic scalar product \eq{eq:scalXY}.) In addition, since we work with real degrees of freedom $q_i$, the spectrum of $\symbHS$ is also invariant under $\la \to -\la^*$ from the invariance of the evolution equation under the (formal) transformation $(q_i, p_i) \to (q_i^*, p_i^*)$. Another way to see this is to note that $\symbHS^* = -\symbHS$. A distinction can then be made between two different types of DIM: 
\begin{itemize}
\item If only two eigenvalues of $\symbHS$ are complex, we speak of a degenerate DIM. In that case the two complex eigenvalues of $\symbHS$ are purely imaginary and opposite to each other. The associated subsystem contains one real degree of freedom: an 
upside-down harmonic oscillator. 
\item If the four eigenvalues of $\symbHS$ are complex, we speak of a nondegenerate DIM. These four eigenvalues are then $\la, \la^*, -\la$, and $-\la^*$. The associated subsystem contains one complex degree of freedom: a rotating 
upside-down complex oscillator, where $\Re \la$ gives the angular velocity, and $\Im \la$ the growth rate.
\end{itemize}
We can now discuss the general relationships between the eigenvalues of $\symbHS$ and those of $\symbHE$. 
\begin{itemize}
\item First, since the energy $H$ of \eq{H1} is conserved in time, a DIM necessarily has a vanishing energy. Thus a dynamical instability requires at least one EI. More precisely, it requires that $\symbHE$ has two eigenvalues with opposite signs. When all the eigenvalues of $M$ have the same sign, there is thus no dynamical instability and all the eigenvalues of $\symbHS$ are real. 
\item One degenerate dynamical instability is either created or erased each time one eigenvalue of $\symbHE$ changes sign 
(this is shown more generally in the next subsection). So, when $\symbHE$ has three strictly positive eigenvalues and one strictly negative one (or conversely), $\symbHS$ has two complex-conjugate, purely imaginary eigenvalues, while its two other eigenvalues remain real. 
\item When a second energetic instability turns in, i.e., when $\symbHE$ has two positive and two negative eigenvalues, either the previous dynamical instability is erased or a new one arises. In the first case, the four eigenvalues of $\symbHS$ are again real. In the second case, they are divided into two sets of complex-conjugate, purely imaginary frequencies $\pm i \Gamma_1$ and $\pm i \Gamma_2$. 
\item Eventually $\Gamma_1$ and $\Gamma_2$ may merge, giving rise to a quartet $\la, \la^*, -\la, -\la^*$ of complex eigenfrequencies. 
\end{itemize}
To illustrate these facts, we consider a matrix $M$ of the form
\be \label{eq:Hqq}
\symbHE=\bmat
 E_{q_1} & \alpha  & 0 & 0 \\
 \alpha  & E_{q_2} & 0 & 0 \\
 0 & 0 & E_{p_1} & 0 \\
 0 & 0 & 0 & E_{p_2}
\emat.
\ee
In \fig{Fig:omBHL}, the evolution of eigenvalues of $M$ and those of $\symbHS$ are shown for increasing values of the off-diagonal term $\alpha$ in the case where the last eigenvalue follows $E_{p_2}=E_{p_2}^0 - \alpha $. 

For $\alpha = 0$ all eigen-values $E_i$ are taken positive. The system is thus stable. At a first critical value of $\alpha$, close to $0.3$ in the present case, an energetic instability appears as one eigenvalue of $M$ becomes negative. Simultaneously, two real eigenvalues of $\symbHS$ merge and become purely imaginary. 
When  $\alpha$ reaches $1.55$, a second  energetic instability appears as another eigenvalue of $M$ becomes negative. At that value, the other two real eigenvalues of $\symbHS$ merge and become purely imaginary. Finally, at a third critical value of $\alpha$ close to $2.16$, the two imaginary eigen-frequencies merge, giving nondegenerate DIM. This third step is not associated as the appearance/disappearance of an EI. 

This simple model thus captures all essential aspects of the ``three-step process'' observed in a black hole laser setup~\cite{MichelParentani} and summarized in App.~\ref{bhli1}. We now understand that this sequence essentially comes from two properties of the mode equation: its symplectic structure and the description using real degrees of freedom. For the Bogoliubov-de Gennes equation, the first property directly comes from the Lagrangian formulation, while the second one can be traced back to the symmetry of the spectrum under $\la \to -\la^*$ (see the last paragraph of the present appendix).  

\subsection{Generalizations}

This analysis can be straightforwardly generalized to a larger number $N$ of degrees of freedom, and to an arbitrary symmetric matrix $\symbHE$. Then, 
\be 
\symbHS = 2 i J \symbHE,
\ee
where 
\be 
J=\left(
\begin{array}{cc}
0 & \textbf{1}_N \\
-\textbf{1}_N & 0
\end{array}
\right).
\ee
As in the previous case, an energetic instability corresponds to a negative eigenvalue of $M$, a degenerate DIM to a pair $(\la, \la^*)$ of purely imaginary eigenvalues of $H_s$, and a nondegenerate DIM to a quartet $(\la, \la^*, -\la, -\la^*)$ of eigenvalues of $H_s$. Moreover, the number of degenerate DIM must change each time a new energetic instability appears since the numbers of DIM and negative-energy modes have the same parity. Indeed, a degenerate DIM corresponds to an imaginary eigenvalue of $\symbHS$ with a positive imaginary part, i.e., to a positive eigenvalue of $J \symbHE$. On the other hand, a negative-energy mode corresponds to a negative eigenvalue of $\symbHE$. Since the characteristic polynomial of $\symbHS$ is of even order $2N$, it as an even number of real roots. Among them, there are an even number of positive roots if $\det \symbHS > 0$ and an odd number of them if $\det \symbHS < 0$, as $\det \lp \symbHS -\lambda \rp$ goes to $+ \infty$ for $\la \to \pm \infty$. Similarly, there are an even number of energetic instabilities if $\det \symbHE > 0$ and an odd number of them if $\det \symbHE < 0$. Noticing that $\det \symbHS = \det \symbHE$ since $\det J = 1$, we obtain the above result. In particular, if there is only one energetic instability, there is one degenerate DIM.~\footnote{An alternative derivation of this statement is to note that only dynamically unstable modes contribute negatively to the energy, see Eq.~(15) in \cite{Coutant10}.} To obtain a nondegenerate DIM requires (at least) a second EI. 

The above naturally extends to complex degrees of freedom, as in the electric model, after decomposing each of them into a pair of real ones. Doing this multiplies the sizes of $\symbHE$ and $\symbHS$ by two. So, two eigenvalues of $H_S$ correspond to a single eigenfrequency for the initial system. To identify these two frequencies, we note that $\symbHS$ expressed in terms of real degrees of freedom has a symmetry under complex conjugation sending  $\la$ to $-\la^*$  which was not present in the initial model. So, two eigenfrequencies $\la$, $-\la^*$ in general correspond to the single eigenfrequency $\la$ for the initial model. The quartet $(\la, \la^*, -\la, -\la^*)$ associated with a nondegenerate DIM thus becomes a doublet $(\la, \la^*)$. A degenerate DIM, which involves only two real degrees of freedom, will not appear in general, except if the modes involved are described by a real degree of freedom. This case never occurs in the electric case discussed in the main text. But it appeared in the case studied in \cite{MichelParentani} for modes with a purely imaginary frequency. 

\section{Spectral theory of the Klein-Gordon equation}
\label{App:KGmath}
\subsection{Resolvent of the Klein-Gordon equation}
\label{sub:Res_KG}
To define a resolvent operator for the Klein-Gordon equation \eqref{KGE}, we rewrite it as a first order in time differential equation. We proceed as in the above Appendix: we switch to Hamiltonian formalism and define the conjugate momentum 
\be
\pi(x,t) \equiv \frac{\delta \mathcal L}{\delta (\partial_t \phi)^*} = (\partial_t + ie A_0(x) ) \phi(x,t).
\ee
Notice that this definition is the complex conjugate of the standard one. We adopt it here as it simplifies the notations. \eq{KGE} is then equivalent to the equation 
\be \label{eq:HamilKG}
i\partial_t \bmat \phi \\ \pi \emat = H_{s} \cdot \begin{pmatrix} \phi \\ \pi \end{pmatrix} = \begin{pmatrix} e A_0(x) & i \\ -i(-\partial_x^2 + m^2) & e A_0(x) \end{pmatrix} \cdot \begin{pmatrix} \phi \\ \pi \end{pmatrix} \label{Hs} .
\ee
where $\phi$ and $\pi$ are treated as independent variables. $H_s$ is an operator acting on phase space and encodes the dynamics. The above equation resembles to a Sch\"odinger equation, but we underline that it is associated with the non-positive definite scalar product of \eq{Ksp}. The resolvent operator is defined in the standard way~\cite{ReedSimon,LevyBruhl} 
\be \label{Res_def}
R_\la = (\la - H_s)^{-1} .
\ee
We shall represent this operator using the $2\times 2$ kernel 
\be
R_\la(x,x') = \bmat A(x,x') & B(x,x') \\ C(x,x') & D(x,x') \emat, 
\ee
where $A$, $B$, $C$, and $D$ are four differential operators such that 
\be
R_\la \cdot \bmat \phi \\ \pi \emat (x) = \bmat \int A(x,x') \phi(x') dx' + \int B(x,x') \pi(x') dx' \\ 
\int C(x,x') \phi(x') dx' + \int D(x,x') \pi(x') dx' \emat. 
\ee
The resolvent allows us to solve the time-dependent equation using a contour integral around the spectrum, see \eq{eq:Res_contour}. 

The legitimacy of \eq{eq:Res_contour} is a highly nontrivial statement in the presence of a Krein product and its precise validity conditions are still unknown. However, it was explicitly proven in the case of the 1+1 dimensional Klein-Gordon equation with a potential~\cite{Bachelot04} or in the case of a ``Pontryagin space'' (i.e., finite number of DIM, as in section~\ref{sec:Model})~\cite{Langer82,Gerard12}. In our case, it is possible to obtain the analytical expression of the resolvent operator. To proceed, we start from the contour integral of \eq{eq:phit} and we identify the resolvent using \eq{eq:Res_contour}. Unfortunately, this determines the resolvent only up to a holomorphic (on the whole complex plane) function. To obtain the full expression, we determine what must be added for the resolvent to satisfy its definition \eqref{Res_def}. 
One can straightforwardly verify that 
\be
R_\la(x,x') = G_\la(x,x') \bmat \la-e A_0(x') & i \\ - i (\la - e A_0(x))(\la - e A_0(x')) + i(\la - e A_0(x'))^2 +i(\partial_{x'}^2 - m^2) & \la-e A_0(x) \emat , \label{res}
\ee
where $G_\la(x,x')$ is the Green function of \eq{Gfunction} satisfies $R_\la \cdot (\la - H_s) = 1$. From \eq{res}, the Green function appears as the ``$2\times 2$ determinant'' of the resolvent. The key implication of this identity is that the resolvent operator $R_\la$ and the Green function $G_\la$ share the same \emph{analytic structure}, i.e., they have the same singularities. This establishes the link between the spectrum and the Green function, and allows us to explicitly relate the Klein-Gordon equation studied in section~\ref{sec:KG} to the Friedrichs model of section~\ref{sec:Model}. 
\medskip

For interested readers, we give below some details that lead us to \eq{res}. The resolvent can be extracted by identifying \eq{eq:phit} and \eq{eq:Res_contour}. A naive guess gives 
\be
R_\la^{\rm naive}(x,x') = G_\la(x,x') \bmat \la-e A_0(x') & i \\ - i (\la - e A_0(x))(\la - e A_0(x')) & \la-e A_0(x) \emat ,
\label{eq:resbis} 
\ee
but it does not satisfy $R_\la^{\rm naive} (\la - H_s) = 1$. However, it is easy to check that the difference with \eqref{res} is analytic in $\la$ 
on the \emph{whole} complex plane since
\be
R_\la^{\rm naive}- R_\la = \bmat 0 & 0 \\  -i\left[(\la - e A_0(x'))^2 +(\partial_{x'}^2 - m^2)\right]G_\la(x,x') & 0 \emat = \bmat 0 & 0 \\  -i \delta(x,x') & 0 \emat.
\ee
This guarantees that the integral \eqref{eq:phit} is unchanged whether one use $R_\la^{\rm naive}$ or $R_\la$. (The first equality can be checked by evaluating the difference on a test function and integrating by parts twice.) 

\subsection{Klein region and dynamical instability}
\label{app:Klein} 

As we saw in the square potential example (section \ref{sub:squarepotential}), a DIM can only appear when $eA > 2m$, so that the maximum potential difference gives enough energy to create a pair of particle/anti-particle. In this appendix, we show that it holds for any smooth potential. The physical idea behind this is clear: a DIM can only occur if the frequency of the emitted particles corresponds to trapped antiparticles (or conversely) in some region of space, allowing the production of pairs at zero energy cost. 

The aim of this subsection is to make this statement more precise and see how it arises from \eq{KGE}. We also show that the real part of a complex eigenfrequency 
always lies between the extremal values of $e A_0$, and give an upper bound on its imaginary part. For definiteness, we assume $e > 0$. Let us consider a complex-frequency mode:
\be 
i \partial_t \phi_\la =\la  \phi_\la,
\ee
where $\la = \om + i \Gamma$ is a complex (nonreal) frequency with real part $\om$ and imaginary part $\Gamma$. We assume that $\phi_\la$ is either $L^2$ 
or periodic in $x$. These are the boundary conditions we used in the body of the text (the first one in the continuous case and the second one in the discrete case). They will allow us to perform integrations by parts without picking additional terms. We also assume that the electrostatic potential has finite lower and upper bounds $V_{\rm min}$ and $V_{\rm max}$: 
\be 
V_{\rm min} \leqslant e A_0(x) \leqslant V_{\rm max}. 
\ee 
In the following, the domain of the integrals is either $\mathbb{R}$ if $\phi_\la$ is $L^2$, or one period if $\phi_\la$ is periodic. In either case, the norm and energy vanish for a DIM. Vanishing of the norm gives 
\be \label{eq:D2}
\int \left(\om- e A_0(x)\right) \left\lvert \phi_\la (t, x) \right\rvert^2 dx=0.
\ee
From this equation, we directly see that $\om \in [V_{\rm min}, V_{\rm max}]$. The vanishing of the energy gives 
\be \label{eq:D1}
\int \left(\left(\om -e A_0(x)\right)^2-\Gamma^2-m^2\right) \left\lvert \phi_\la (t, x) \right\rvert^2 dx
\geq 0.
\ee
(It can also be obtained from the equality $\int \pd_x^2 \left\lvert \phi_\om \right\rvert^2 dx = 0$ after a few lines of algebra.) 

To proceed, it is convenient to write the integrand in terms of $V_{\rm eff}(x) \equiv \lp \om - e A(x) \rp \lp V_{\rm min} + V_{\rm max} - 2 e A_0(x) \rp$. 
Intuitively, the sign of the first factor distinguishes between what can locally be called positive or negative charged particle. The second factor is the deviation of the local potential with respect to 
the median one $\lp V_{\rm min} + V_{\rm max}\rp/2$. Their product can thus be seen as an effective (shifted) potential, counted positively for particles and negatively for antiparticles. To arrive at this form, we multiply \eq{eq:D2} by $2 \om - V_{\rm min} - V_{\rm max}$ and subtract it to twice \eq{eq:D1}: 
\be \label{bbb}
\int  (V_{\rm eff}(x)-2\Gamma^2-2m^2) \left\lvert \phi_\la (t, x) \right\rvert^2 \geq 0.
\ee
The first term is the aforementioned effective potential, which will be our main tool in the derivation. The two other terms give a shift due to the imaginary part of $\la$ and to the mass. \eq{bbb} may be interpreted in the following way. The effective potential $V_{\rm eff}$ is the driving force for the mode amplification, acting effectively as an attractive term which makes pair production more favourable the more pairs are already present. The mass term, on the other hand, makes pair production less favourable. A self-amplified behaviour can thus arise only if the overlap of $V_{\rm eff}$ and $\phi_\la$ is large enough to make pair production energetically favourable despite the mass barrier. Using the inequality
\be 
V_{\rm eff}(x) \leq |\om - e A_0(x)| \lp V_{\rm max} - V_{\rm min} \rp, 
\ee
we get  
\be 
\int V_{\rm eff}(x) \left\lvert \phi_\la (t, x) \right\rvert^2dx \leq  \left(V_{\rm max}-V_{\rm min} \right)\int | \om-e A_0(x) | \, \left\lvert \phi_\la (t, x) \right\rvert^2dx . 
\ee
To go further, it is useful to separate this quantity into two integrals whose domains are determined by the sign of $\omega - e A_0$. 
We denote as $L_+$ the domain on which it is positive, and $L_-$ the domain on which it is negative. We obtain 
\be \label{aaa} 
\int V_{\rm eff}(x) \left\lvert \phi_\la (t, x) \right\rvert^2dx \hspace*{-0.27 cm} & \leq & \hspace*{-0.27 cm} \left(V_{\rm max}-V_{\rm min}\right)\int_{L_+}  \lp \om-e A_0(x) \rp \, \left\lvert \phi_\la (t, x) \right\rvert^2dx \nn
& & + \left(V_{\rm max}-V_{\rm min}\right)\int_{L_-}  \lp e A_0(x) - \om \rp \, \left\lvert \phi_\la (t, x) \right\rvert^2 dx.
\ee 
Moreover, from Eq.~(\ref{eq:D2}) the two terms in the right-hand side are equal, so 
\be \label{ineq1}
\int V_{\rm eff}(x) \left\lvert \phi_\la (t, x) \right\rvert^2dx 
\leq 2 \left(V_{\rm max}-V_{\rm min}\right)\int_{L_+}  \lp \om-e A_0(x) \rp \, \left\lvert \phi_\la (t, x) \right\rvert^2dx 
\ee
and
\be \label{ineq2}
\int V_{\rm eff}(x) \left\lvert \phi_\la (t, x) \right\rvert^2dx 
\leq 2 \left(V_{\rm max}-V_{\rm min}\right)\int_{L_-}  \lp e A_0(x) - \om \rp \, \left\lvert \phi_\la (t, x) \right\rvert^2 dx.
\ee
One can now give separate bounds on the right-hand sides of Eqs.~(\ref{ineq1},\ref{ineq2}) using $V_{\rm max} \geq e A_0(x) \geq V_{\rm min}$.
\be \label{eq:longineq} 
\int  V_{\rm eff}(x) \left\lvert \phi_\la (t, x) \right\rvert^2 dx \leq \left(V_{\rm max}-V_{\rm min}\right) \text{Min}\left(2\left(\om - V_{\rm min}\right)\text{  }\int _{L_+} \left\lvert \phi_\la (t, x) \right\rvert^2 dx,2\left(V_{\rm max}-\om \right)\int _{L_-}  \left\lvert \phi_\la (t, x) \right\rvert^2 dx)\right) \nn
\ee
A straightforward calculation shows that for any set of four positive numbers $a_1, a_2, a_3, a_4$,
\be 
\min \lp a_1 a_2, a_3 a_4 \rp \leq \frac{1}{4} \lp a_1 + a_3 \rp \lp a_2 + a_4 \rp.
\ee
Applying this inequality with
\be 
a_1 \equiv 2 \lp \om - V_{\rm min} \rp, \; a_2 \equiv \int _{L_+} \left\lvert \phi_\la (t, x) \right\rvert^2 dx, \; a_3 \equiv 2 \lp V_{\rm max} - \om \rp, \; \text{and} \; a_4 \equiv \int _{L_-} \left\lvert \phi_\la (t, x) \right\rvert^2 dx,
\ee
we find the right-hand side of \eq{eq:longineq} is smaller than
\be 
\frac{1}{2}\left(V_{\rm max}-V_{\rm min}\right)^2 \int \left\lvert \phi_\la (t, x) \right\rvert^2 dx.
\ee
On the other hand, from Eq.~(\ref{bbb}) it must be larger than 
\be 
2 \left(\Gamma^2+m^2\right) \int \left\lvert \phi_\la (t, x) \right\rvert^2 dx.
\ee
Hence we get 
\be \label{resD}
V_{\rm max}-V_{\rm min}\geq 2\sqrt{m^2+\Gamma^2}.
\ee
This is a sufficient condition for the existence of a Klein region, i.e. $V_{\rm max}-V_{\rm min} \geq 2 m$. Note that we also obtain the following upper bound on the growth rate of the instability: 
\be \label{eq_boundGamma}
\Gamma^2 \leq \lp \frac{V_{\rm max}-V_{\rm min}}{2} \rp^2 - m^2.
\ee

\section{Klein regions and black holes}
\label{BH_App}

As mentioned in the Introduction, there exists a deep analogy between the electrostatic model of subsection~\ref{sub:modebasis} and the propagation of a scalar field in curved space-time. This analogy was studied in details in the second part of the Appendix in \cite{Fulling}. In particular, it was shown that a Klein region can serve as a toy-model for the Kerr black hole ergosphere, shedding a new light on superradiance in black hole physics. A more complicate version of this toy-model, taking into account specific features of the Kerr metric, was studied in \cite{Kaloper}. 
In order to show the relevance of the general concepts and results of the present work for black hole physics, let us rephrase the conditions for superradiance and DIM formulated in \cite{Coutant10}, and show how they apply here.

From a semiclassical analysis, it is found that superradiance occurs whenever the two following conditions are met:
\begin{itemize}
\item For some real values of the frequency $\la$, 
the field equation admits both positive-norm and negative-norm WKB modes; 
\item These modes are mixed when considering global exact solutions of the mode equation.
\end{itemize}   
The second condition is a very weak one, as two WKB modes with the same frequency in a stationary, non-uniform background are generally not exact solutions of the mode equation and will be mixed. So the important condition is the first one. Its interpretation in terms of the analogue hydrodynamic metric of~\cite{unruhprl81, Coutant10} for long-wavelength modes is the transonic character of the flow, i.e., the presence of an acoustic horizon. In the electrostatic model of subsection~\ref{sub:modebasis}, it is equivalent to the presence of a Klein region. So, three \emph{a priori} different notions have the very same mathematical origin: superradiance in Kerr black holes, sonic horizons in hydrodynamics, and Klein regions in the electrostatic model. 

Two additional conditions are required for having dynamical instabilities: 
\begin{itemize}
\item One of the aforementioned WKB solutions must be trapped;
\item The depth of the trapping potential must be large enough so that at least one pair of bound modes exist.
\end{itemize}
These additional conditions are satisfied in the toy-model of~\cite{Kaloper} in a large domain of parameter space, leading to the black hole instability. They are also satisfied in the electrostatic or hydrodynamic cases if the Klein or supersonic region, respectively, are finite but large enough~\cite{Fulling, Finazzi10}, hence the laser effect.  

\bibliography{biblio}

\end{document}